\documentclass[12pt]{article}
\usepackage{amsmath}
\usepackage{graphicx}
\usepackage{pdflscape}
\usepackage{enumerate}
\usepackage{color}
\usepackage{graphicx, subfigure}
\usepackage{stackengine}
\usepackage{listings}

\def\refhg{\hangindent=20pt\hangafter=1}
\def\refmark{\par\vskip 2mm\noindent\refhg}
\def\bse{\begin{eqnarray*}}
\def\ese{\end{eqnarray*}}
\begin{document}
\font\fmain=cmbx10 scaled\magstep4
\title{Parametric versus Semi and Nonparametric Regression Models}
\author{Hamdy F. F. Mahmoud$^{1,2}$ }
\date{}
\maketitle
\thispagestyle{empty}
\baselineskip=16pt
\begin{center}
\noindent {$^1$} Department of Statistics, Virginia Tech, Blacksburg VA, USA.\\
\noindent {$^2$} Department of Statistics, Mathematics and Insurance, Assiut University, Egypt.\\
\end{center}



\begin{abstract}
Three types of regression models researchers need to be familiar with and know the requirements of each: parametric, semiparametric and nonparametric regression models. The type of modeling used is based on how much information are available about the form of the relationship between response variable and explanatory variables, and the random error distribution. In this article, differences between models, common methods of estimation, robust estimation, and applications are introduced. The R code for all the graphs and analyses presented here, in this article, is available in the Appendix.

\vskip 5mm \noindent
\underline{\hbox{\bf Keywords}}: Parametric, semiparametric, nonparametric, single index model, robust estimation, kernel regression, spline smoothing. 
\thispagestyle{empty}
\end{abstract}
\section{Introduction}\label{sec:sec1}
The aim of this paper is to answer many questions researchers may have when they fit their data, such as what are the differences between parametric, semi/nonparamertic models? which one should be used to model a real data set? what estimation method should be used?, and which type of modeling is better? These questions and others are addressed by examples.

Assume that a researcher collected data about Y, a response variable, and  explanatory variables, $X=(x_1, x_2, \ldots, x_k)$. The model that describes the relationship between Y and X can be written as
\begin{equation}
Y=f(X,\boldsymbol\beta)+\epsilon,
\end{equation}
where $\boldsymbol\beta$ is a vector of $k$ parameters, $\epsilon$ is an error term whose distribution may or may not be normal, $f(\cdot)$ is some function that describe the relationship between Y and X. 

The model choice from parametric, semiparametric and nonparametric regression model depends on the prior knowledge of the functional form relationship, $f(\cdot)$, and the random error distribution. If the form is known and it is correct, the parametric method can model the data set well. However, if a wrong functional form is chosen, this will result in larger bias as compared to the other compititve models (Fan and Yao, 2003). The most common functional form is parametric linear model, as a type of parametric regression, is frequently used to describe the relationship between a dependent variable and explanatory variables. Parametric linear models require the estimation of a finite number of parameters, $\boldsymbol\beta$. 

Parametric models are easy to work with, estimate, and interpret. So, why are semiparametric and nonparametric regression important? Over the last decade, increasing attention has been devoted to these regression models as new techniques for estimation and forecasting in different areas, such as epidemiology, agriculture, and economics. Nonparametric regression analysis relaxes the assumption of the linearity or even knowledge about the functional form priori in the regression analysis and enables one to explore the data more flexibly. However, in high dimensions variance of the estimates rapidly increases, called as curse of dimensionality, due to the sparseness of data. To solve this problem, some semiparametric methods have been proposed, such as single index model (SIM). In application, the three modeling approaches are compared in terms of fitting and prediction (Rajarathinan and Parameter 2011; Dhekale et al. 2017) and Mahmoud et al. (2016) showed that if the link function in generalized linear models is misspecified, semiparametric models are better than parametric models. In Section 2, the three types of models are introduced. In Section 3, which model should be used is addressed. In Section 4, common estimation methods for semiparametric and nonprametric models are displayed. In Section 5, multiple case is presented. Robust nonparametric regression method is introduced in Section 6. Section 7 is discussion and conclusion.

\section{Parametric, semi and nonparametric regression models}
To differentiate between the three types of regression models, without less of generality, assume we have a response variable, Y, and two explanatory variables, $x_1$ and $x_2$. The regression model that describes the relationship between the response variable and two the explanatory variables can be written as:
\begin{equation}
Y=f_1(x_1,\beta_1)+f_2(x_2,\beta_2)+\epsilon,
\end{equation}
where $\beta_1$ and $\beta_2$ are some parameters need to be estimated, $\epsilon$ is a random error term, and $f_1(\cdot)$  and $f_2(\cdot)$ are functions describe the relationships.
\subsection{Parametric models}
Parametric models are models in which the vector of parameters, $\boldsymbol\beta$ in model (1), is a vector in finite $p-$dimensional space ($p$ may be less or greater than $k$). Our interest in this case is estimating the vector of parameters. In parametric models, the researcher assumes completely the form of the model and its assumptions. Applying on model (2), $f_1(\cdot)$  and $f_2(\cdot)$ need to be known functions and $\epsilon$  distribution is known periori for inference purposes. For example, the researcher may assume they are linear and the error term follow normal distribution. In general,  $f_1(\cdot)$  and $f_2(\cdot)$ can be assumed linear in parameters or nonlinear in the parameters $\beta$'s. For model validity, after fitting the data by the assumed model, the researcher needs to check whether the assumptions are correct using residuals.

\subsubsection{Linear models}
Linear models are linear in the parameters (i.e., linear in the $\beta$'s). For example, polynomial regression that is used to model curvature in a data set by using higher-ordered values of the predictors is a linear model in $\beta$'s. However, the final regression model is just a linear combination of higher-ordered predictors. There are many examples of linear models, below are a few of them
\begin{itemize}
\item $Y=\beta_0+\beta_1 x_1+\beta_2 x_2+\epsilon$   (Multiple linear regression model)
\item $Y=\beta_0+\beta_{10} x_1+\beta_{11} x_1^2+\beta_{20} x_2+\beta_{21} x_2^2+\epsilon$  (Polynomial regression model of second order)
\item $\log (\mu)=\beta_0+\beta_1 x_1+\beta_2 x_2 $  (Poisson regression when Y is count data)
\end{itemize}
These models are linear in the parameters and the number of parameters is greater that the number of the explanatory variables with one ($p=k+1$). For this setting, there are many estimation methods for parameter estimation, such an ordinary least square method and maximum likelihood method. The main interest of the researcher, in this case, is estimating the vector of parameters. Once he estimated the parameters, everything is easy afterwards.

\subsubsection{Nonlinear models}
The nonlinear regression models are parametric models and the function $f(\cdot)$ is known but nonlinear in the parameters. Below are some examples
\begin{itemize}
\item $Y=\beta_0+\beta_1 x_1+\beta_2 e^{(\beta_3 x_2)} +\epsilon$ 
\item $Y=\frac{e^{\beta_0+\beta_1x_1}}{1+e^{\beta_0+\beta_1x_1}}+\epsilon$
\item $Y=\beta_0+(0.4-\beta_0) e^{-\beta_1(x_1-5)}+\beta_2x_2+\epsilon$
\end{itemize}
For these setting, nonlinear least square method can be used to estimate the model parameters. There are many algorithms for nonlinear least squares estimation, such as
\begin{enumerate}
\item Newton's method which is based on a gradient approach but it can be computationally challenging and dependents on initial values.
\item Gauss-Newton algorithm, it is a modification of Newton's method that gives an approximation of the solution that Newton's method but it is not guaranteed to converge.
\item Levenberg-Marquardt method, it has less computational difficulties compared to the other methods but it requires much work to reach the optimal solution.
\end{enumerate}
There are some nonlinear models that can be made linear in the parameters by a simple transformation. For example:
\begin{center}
$Y=\frac{\beta_0x}{\beta_1+x}$
\end{center}
can be written as
\begin{equation}
\frac{1}{Y}=\frac{1}{\beta_0}+\frac{\beta_1}{\beta_0}\frac{1}{x}
\end{equation}

\begin{equation}
Y^*=\theta_0+\theta_1 x^*
\end{equation}
The last equation is linear in the parameters, so it is a linear model.
If the researcher found that the relationship between the response and explanatory variable is not linear, simple solutions may work to fix the problem before working with nonlinear, nonparametric or semiparametric modeling, such as
\begin{itemize}
\item Use a nonlinear transformation to the dependent and/or independent variables such as a log transformation, square root, or power transformation.
\item Add another regressor which is a nonlinear function of one of the other variables. For example, if you have regressed Y on X, it may make sense to regress Y on both X and $X^2$ (i.e., X-squared). 
\end{itemize}
In real application, parametric models do not work for fitting data in many applications (Rajarathinan and Parmar (2011).

\subsection{Nonparametric models}
The parametric and nonparametric models differ in that the model form is not specified a priori but is instead determined from data set. The term nonparametric does not mean that such models are completely lack parameters, but that the number of the parameters are flexible and not fixed periori. In parametric models, the vector of parameter, $\boldsymbol\beta$, is a vector in finite $p-$ dimensional space and our main interest is estimating that vector of parameters. In contrast, in nonparametric models, the set of parameters is a subset of infinite dimensional vector spaces. The primary interest is in estimating that infinite-dimensional vector of parameters. In nonparametric regression models, the relationship between the explanatory variables and response is unknown. Applying on model (2), $f_1$ and $f_2$ both are unknown functions. These functions can take any shape but they are unknown to the researcher, they maybe linear or nonlinear relationship but they are unknown to the researcher.
In parametric modeling, you know which model exactly you use to fit to the data, e.g., linear regression line. The data tells you what the regression model should look like;  the data will decide what the functions, $f_1$ and $f_2$, looks like
\begin{figure}[h!]
\footnotesize
\hspace{0.5cm}\stackunder[5pt]{\includegraphics[width=7.5cm,height=6cm]{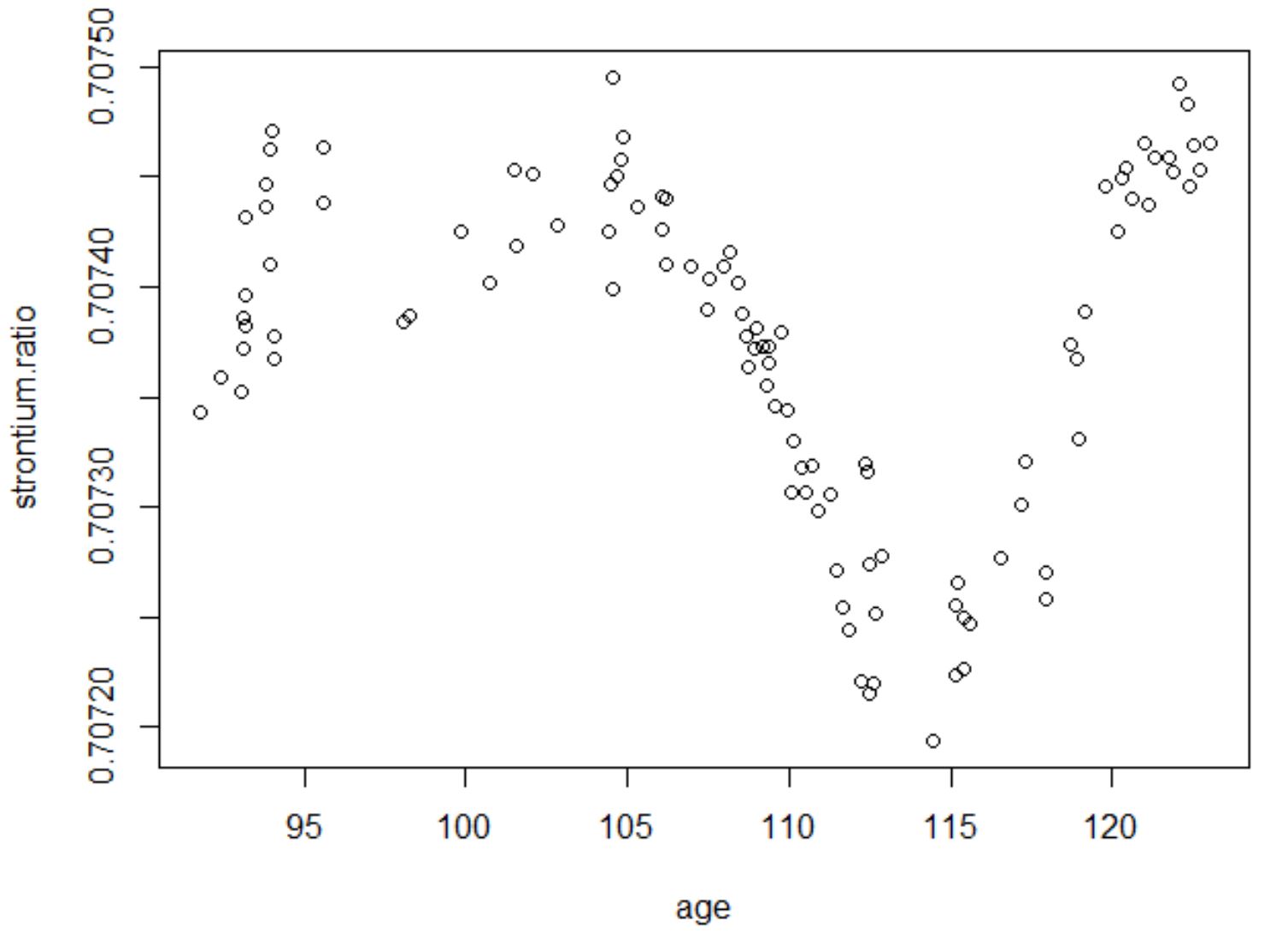}}{(a)}
\stackunder[5pt]{\includegraphics[width=8cm,height=6cm]{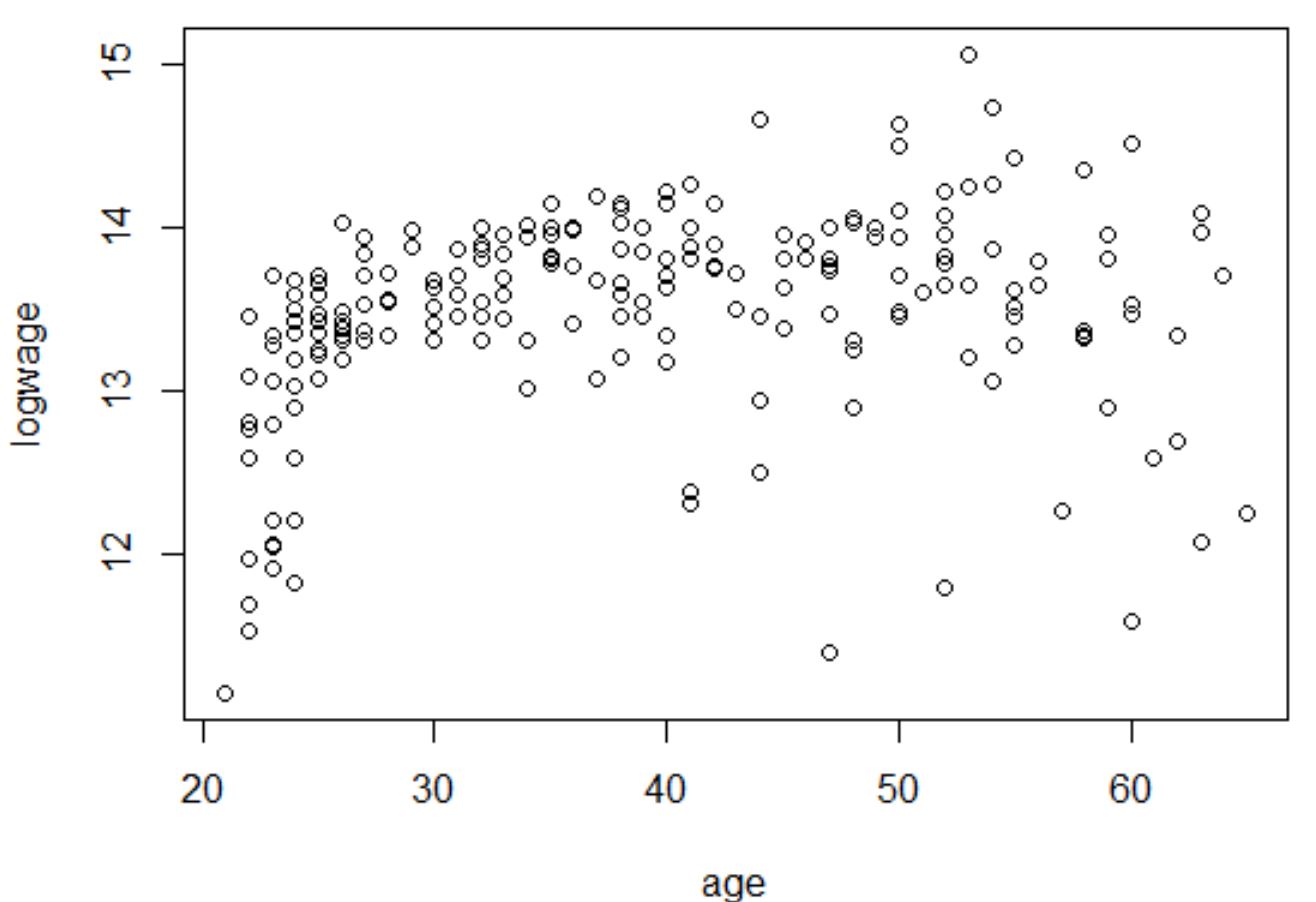}}{(b)}
\stackunder[5pt]{\includegraphics[width=8cm,height=6cm]{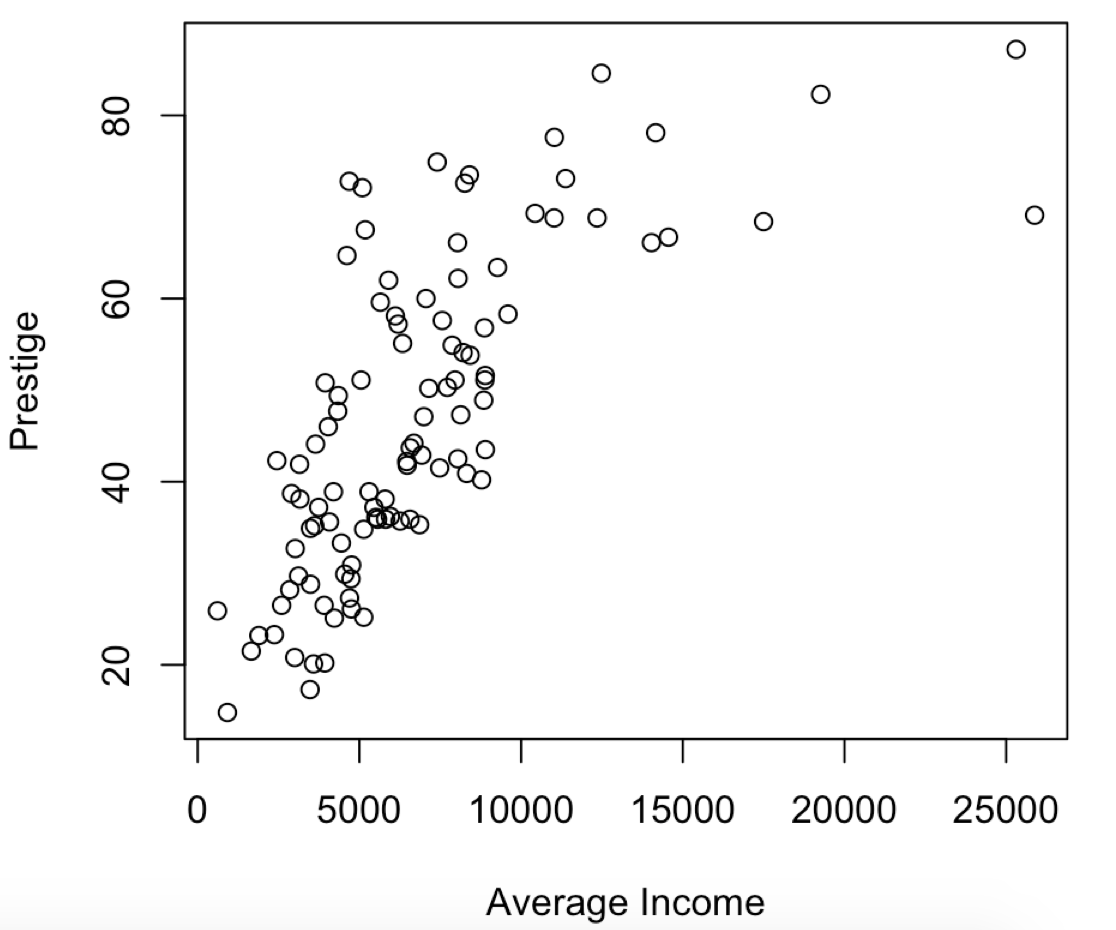}}{(c)}
\hspace{0.5cm}%
\stackunder[5pt]{\includegraphics[width=8cm,height=6cm]{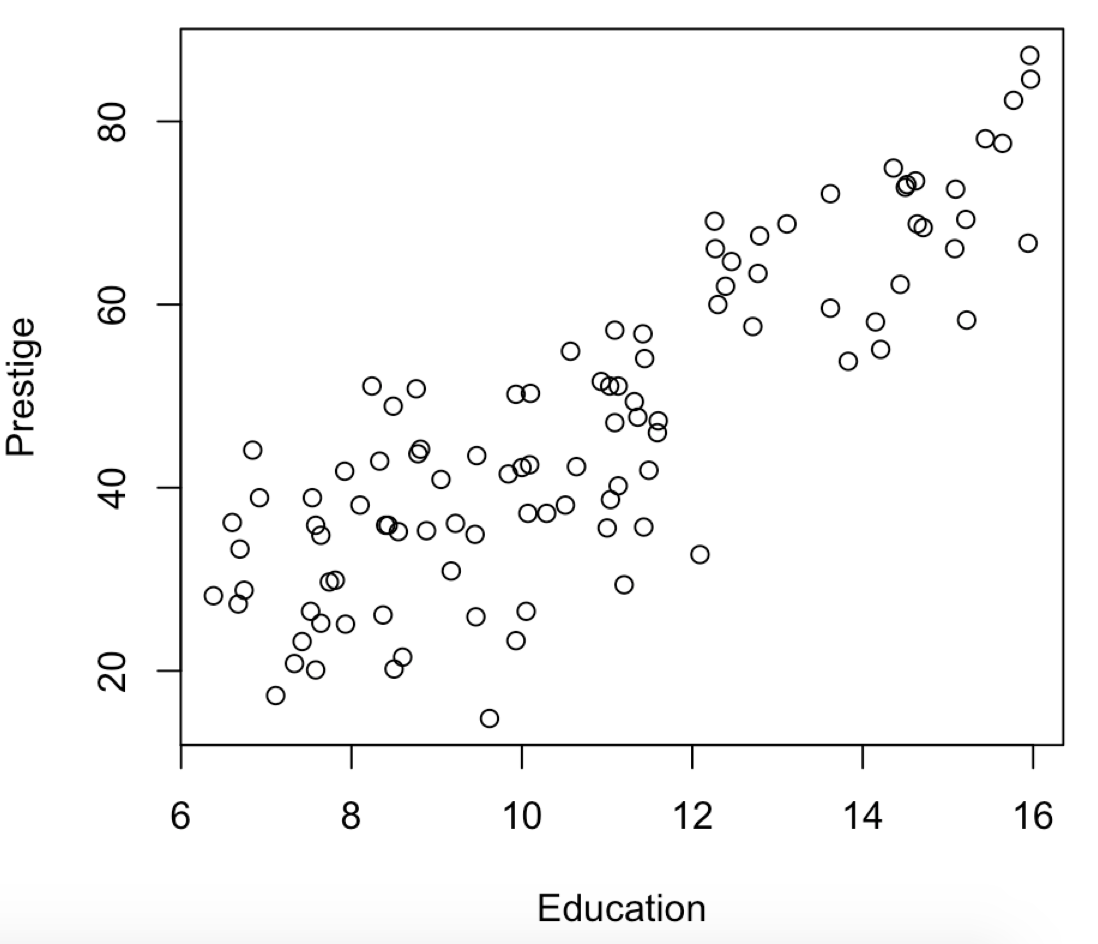}}{(d)}
\caption{A scatter plot of age and strontium ratio (a), age versus log of wage (b), income and prestige (c), and education and prestige (d).}
\end{figure}

\subsection{Semiparametric models}
Semiparametric modeling is a hybrid of the parametric and nonparametric approaches of statistical models. 
It may appear at first that semiparametric model include nonparametric model, however, semiparametric model is considered to be "smaller" than a completely nonparametric model because we are often interested only in the finite-dimensional component of $\boldsymbol\beta$. By contrast, in nonparametric models, the primary interest is in estimating the infinite-dimensional parameter. In result, the estimation is statistically harder in nonparametric models compared to semiparametric models.

While parametric models are being easy to understand and easy to work with, they fail to give a fair representation of what is happening in the real world. Semiparametric models allow you to have the best of both worlds: a model that is understandable and offering a fair representation of the messiness that is involved in real life.
Semiparametric regression models take many different structures. One is a form of regression analysis in which a part of the predictors do not take predetermined forms and the other part takes known forms with the response. For example, in model (2), $f_1$ may be known (assume linear) and $f_2$ is unknown, it can be written as
\begin{center}
$Y=\beta_0+\beta_1 x_1+f(x_2)+\epsilon$
\end{center}
In this setting, the relationship between $x_1$ and the response is linear but the relationship between the response and $x_2$ is unknown.
Another form of semiparametric regression modeling which is a well-known example is the single index model (SIM) that is extensively studied and has many applications. In general, it takes the form
\begin{equation}
Y=f(X\boldsymbol\beta)+\epsilon,
\end{equation}
where $f$ is an unknown function, $X = (x_1, \ldots, x_k)$ is a $n \times k$ matrix of regressors values, $\boldsymbol\beta$ is a $k \times 1$ vector of parameters, and and $\epsilon$ is the error satisfying E$(\epsilon|X) = 0$. The term $X \boldsymbol\beta$ is called a ``single index" because it is a scalar (a single index). The functional form of $f(\cdot)$ is unknown to the researcher. This model is semiparametric since the functional form of the linear index is specified, while $f(\cdot)$ is unspecified. SIM is extensively studied and applied in many different fields including biostatistics, medicine, economics, financial econometrics and epidemology (Ruppert et al. 2003; Mahmoud et. al 2016, 2019; Toma and Fulga 2018;  Li, et al. 2017; Qin et al. 2018).

SIM is more flexible compared to parametric models and does not lack from curse of dimentionality problem compared to nonparametric models. It assumes the link between the mean response and the explanatory variables is unknown and estimates it nonparametrically. This gives the single index model two main advantages over parametric and nonparametric models: (1) It avoids mispecifying the link function and its misleading results (Horowitz and Hardle, 1996) and (2) the reduction of dimension which is achieved by assuming the link function to be a univariate function applied to the projection of explanatory covariate vector on to some direction. For estimation, the coefficient of one component of X is set to be equal to one and needs to be continuous to fix the identifiability problem (Ichimura 1993, Sherman 1994) or to set $\boldsymbol\beta$=1 (Lin and Kulasekera 2007, Xia et al. 2004). 

Model (2) can be written in the form of single index model as 
\begin{equation}
Y=f(\beta_1 x_1+\beta_2 x_2)+\epsilon
\end{equation}
Figure 1 shows different relationships between a response variable and an explanatory variable. For Figure 1(a) - 1(c), the relationship between the response and explanatory variable is not linear, so a researcher may use nonparamteric model to fit this data or try to find a polynomial regression model that can fit the data because there is no known model for these data sets. Figure 1(d) shows a linear relationship between the two variables.

\section{Which type of modeling you should use to fit your data?}
The first step in statistical analysis is summarizing the data numerically and graphically for the response (dependent) variable and the explanatory (independent) variables. The researcher should look at scatter plots, boxplots, histograms, numerical summery and others. That is to get an initial knowledge about the data and see whether the model assumptions are satisfied and whether there are outliers.

Figure 1 shows the relationship between age and log of wage is nonlinear. Figure 2 display the fitted lines of linear, quadratic, cubic and the 4th degree of polynomial along with the scatterplot.

\begin{figure}[h!]
\includegraphics[width=16.5cm, height=12cm]{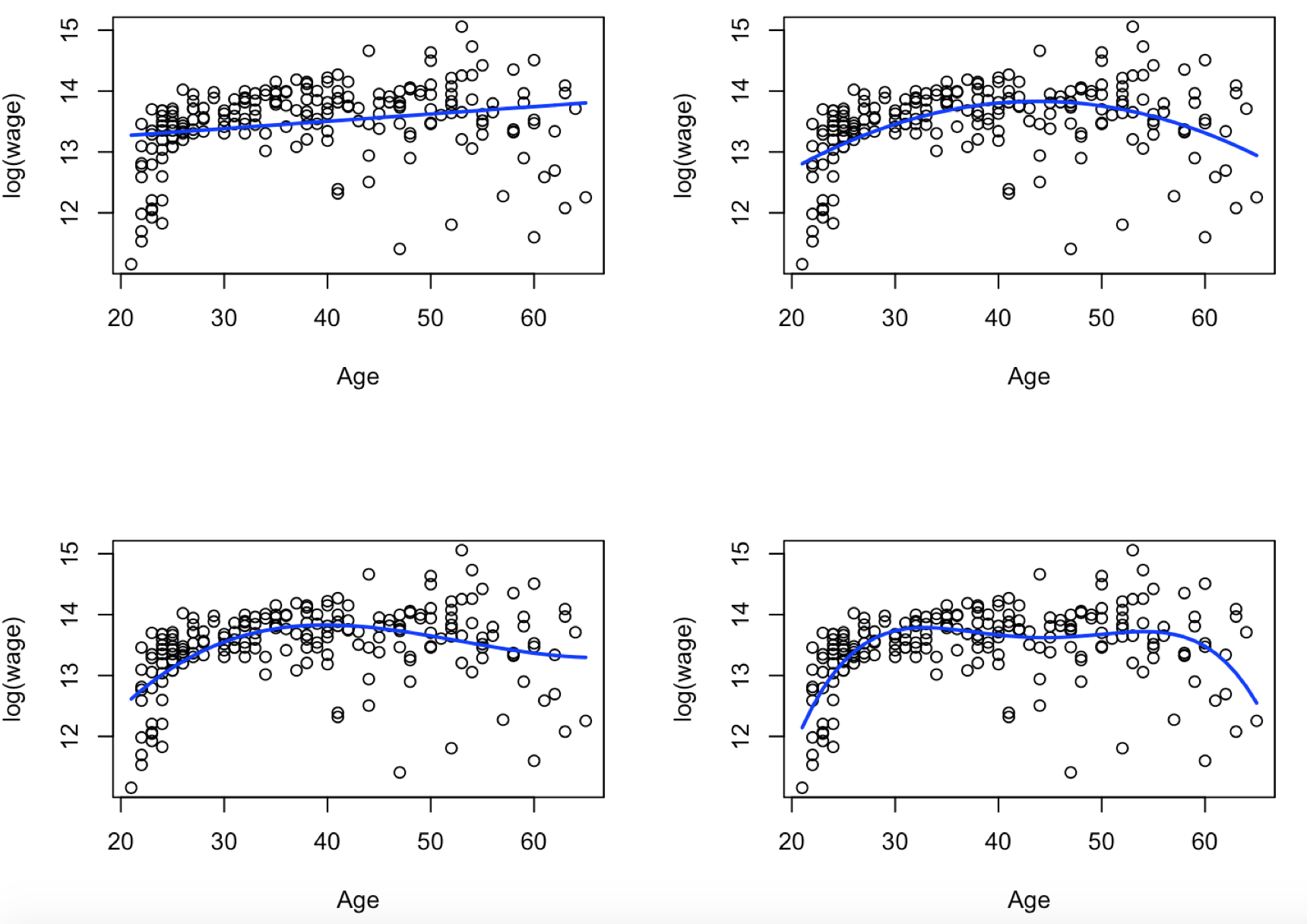}
\caption{A scatterplot of age and log(wage) along with different polynomial regression lines: linear $y=\beta_0+\beta_1x$ (top left), quadratic $y=\beta_0+\beta_1x+\beta_2x^2$ (top right), cubic $y=\beta_0+\beta_1x+\beta_2x^2+\beta_3x^3$ (bottom left), and quartic $y=\beta_0+\beta_1x+\beta_2x^2+\beta_3x^3+\beta_4x^4$ (bottom right).}
\end{figure}

\begin{figure}[h!]
\includegraphics[width=15.5cm, height=10.3cm]{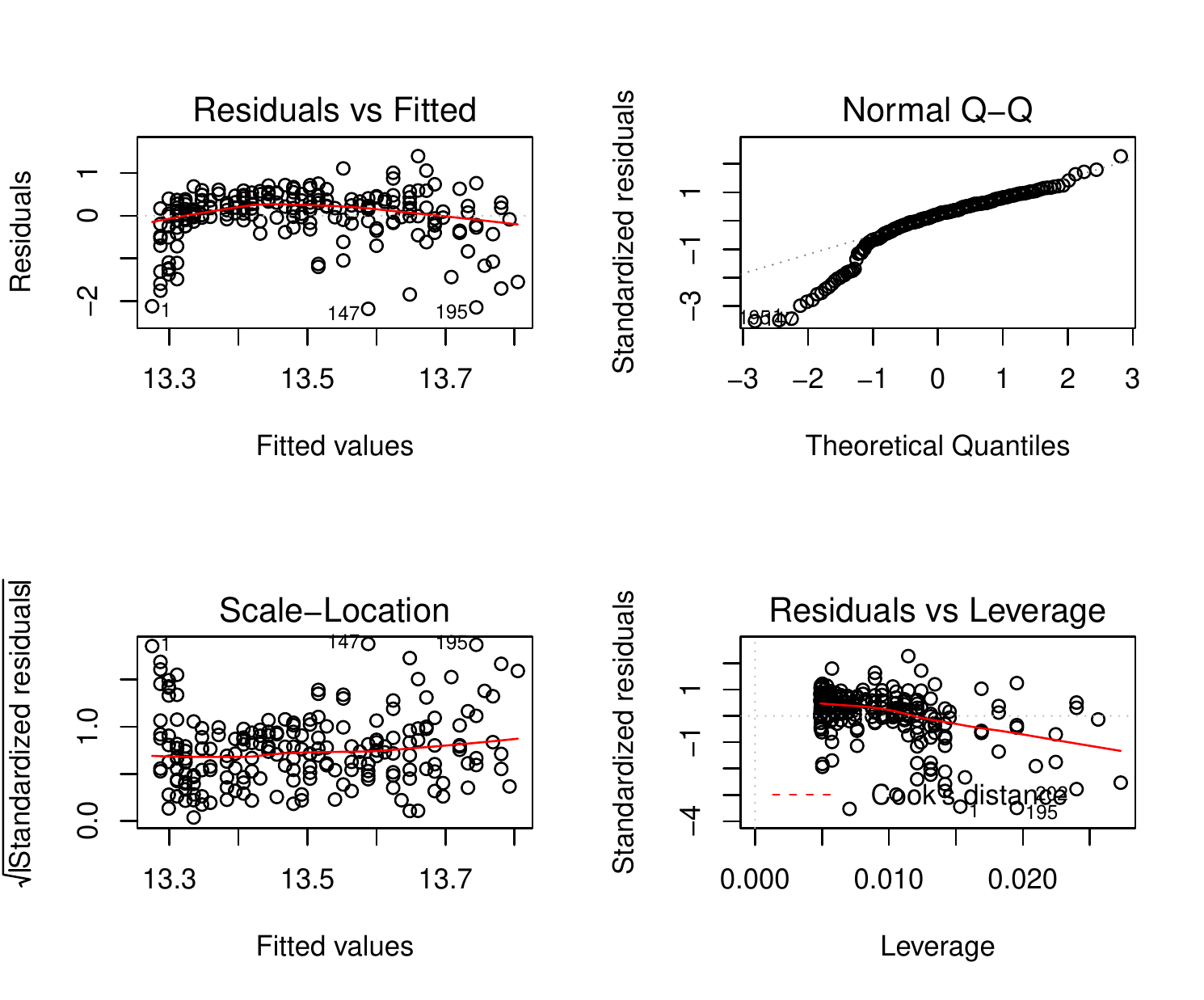}
\vskip -5mm
\caption{Residual plots for the linear model, $y=\beta_0+\beta_1x$.}
\end{figure}
\begin{figure}[h!]
\includegraphics[width=15.5cm, height=10.3cm]{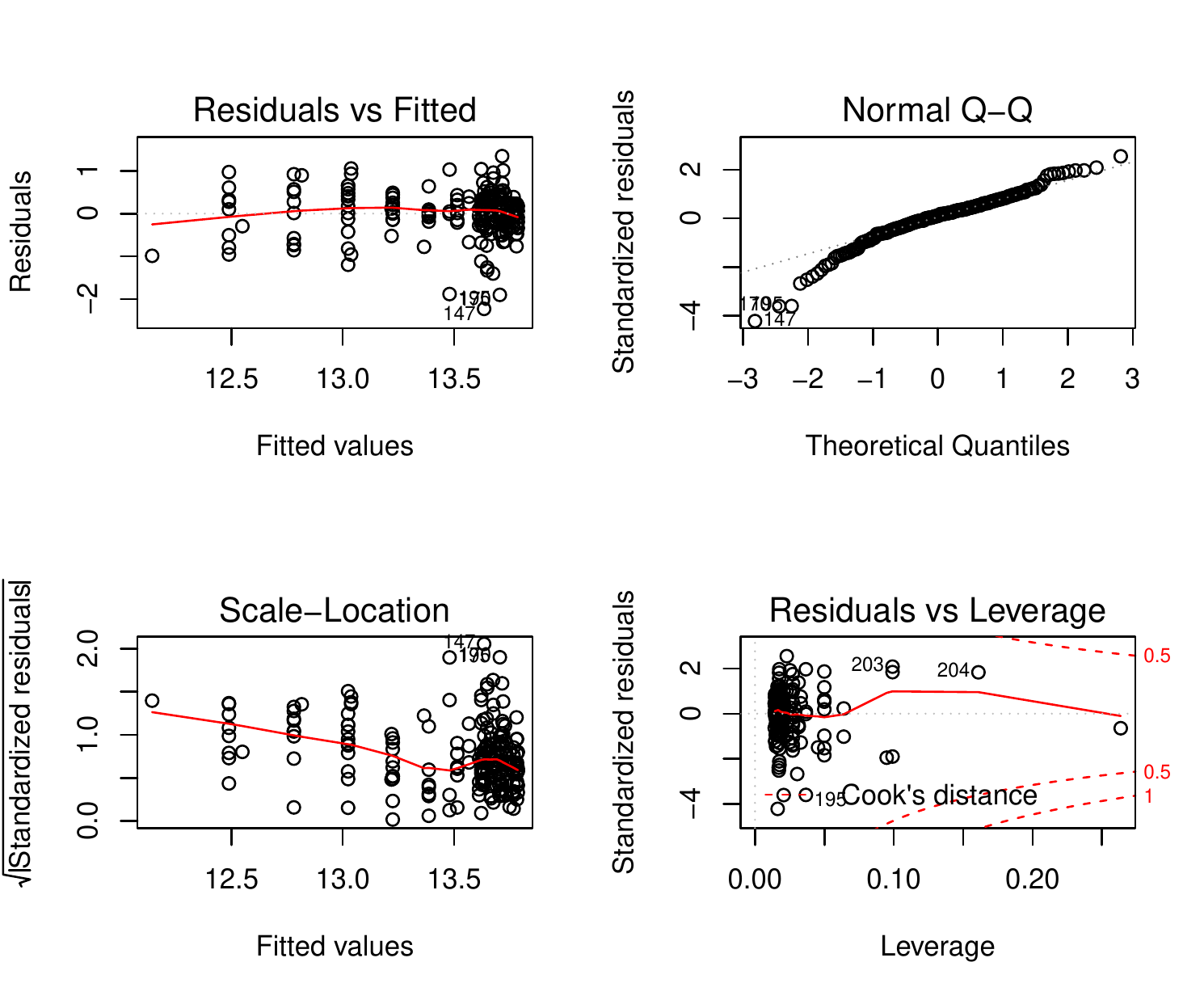}
\vskip -5mm
\caption{Residual plots for the $4^{th}$ degree polynomial model, $y=\beta_0+\beta_1x+\beta_2x^2+\beta_3x^3+\beta_4x^4$.}
\end{figure}

Assume that a researcher used the linear model to fit data. To check linearity, the researcher needs to look at residual plots. Figure 3 shows that linearity is not satisfied because fitted values versus residuals reveals nonlinearlty form and also Normality is not satisfied from Normal Q-Q plot. Compared the linear model to $4^{th}$ degree polynomial model, Figure 4 shows a better fitted versus residuals graph and normality of residual is much better. In addition, we need to look at p-value associated with the fitted model. For the linear model, p-value = 0.0008407 which is significant, however it is clear that linear model is not suitable to fit these data and R-squared = 0.05357 which is very small. For the 4th degree polynomial regression, p-value = 1.202e-15, R-squared = 0.315. Now we need to cover semi and nonparametric regression models. 

\section{Common estimation methods of semi and nonparamteric regression models}
Estimating the unknown function nonparametrically means the data itself is used to estimate the function, $f(\cdot)$, that describes the relationship between explanatory variables and a response. There are two commonly used approaches to nonparametric regression term:
\begin{enumerate}  
\item {\bf Kernel Regression:} estimates the conditional expectation of Y at given value $x$ using a weighted filter to the data.
\item {\bf Spline Smoothing:} minimize the sum of squared residuals plus a term which penalizes the roughness of the fit.
\end{enumerate}
\subsection{Kernel Regression} 

One of the most popular methods for nonparametric kernel regression was proposed by Nadaraya (1965) and Watson (1964) and is known as the Nadaraya-Watson estimator (also known as the local constant estimator), though the local polynomial estimator has emerged as a popular alternative. For more details in kernel smoothing, see Wand and Jones (1995).

The Kernel regression estimation is to estimate the mean function at a given value of $x$ with a polynomial with order $p$ using a kernel function that assign high weight to the point $x$ and proportional weight to the other values based on the distance. To estimate $\hat{f}(x)$ at a given value $x$ as a locally weighted average of all $y's$ associated to the values around $x$, the Nadaraya-Watson estimator is:
\begin{equation}
 \widehat{f}_h(x)=\frac{\sum_{i=1}^n K(\frac{x-x_i}{h}) y_i}{\sum_{i=1}^nK(\frac{x-x_i}{h})}
\end{equation}
where K is a Kernel function (weight function) with a bandwidth $h$. $K$ function should give weights decline as one moves away from the target value.
Local polynomial estimate, $\boldsymbol\beta^T=(\beta_0, \beta_1, \ldots, \beta_p)$, is obtained by minimizing the following quantity:
\begin{equation}
\Sigma_{i=1}^{n}K_h(x_i-x)\{Y_i-[\beta_0+\beta_1(x_i-x)+\ldots, \frac{1}{j!} \beta_p(x_i-x)^p]\}^2,
\end{equation} 
where K is a kernel function and $h$ is a bandwidth controlling the degree of smoothing. When $p=0$, the local polynomial is a local constant. Loader (1999) explained cross validation (CV), $CV(h)=\sum_{i=1}^{n}[y_i-\hat{f_{h,-i}}(x_i)]^2$, and generalized cross-validation (GCV) are criteria for bandwidth selection in local polynomial. Also there is a rule of thumb in case Gaussian function is used then it can be shown that the optimal choice for $h$ is
\begin{center}
$h = \left(\frac{4\hat{\sigma}^5}{3n}\right)^{\frac{1}{5}} \approx 1.06 \hat{\sigma} n^{-1/5}$,
\end{center}
where $\hat{\sigma}$ is the standard deviation of the samples and $n$ is the sample size. Below are some common kernel functions 
\begin{itemize}
\item Epanechnikov: $K(\cdot)= \frac{3}{4}(1-d^2)$, $d^2 <$  1, 0 otherwise,
\item Minimum var:  $K(\cdot)=\frac{3}{8}(1-5d^2)$, $d^2<$  1, 0 otherwise,
\item Gaussian density: $exp(-\frac{x-x_i}{h})$
\item Tricube function: $W(z)=(1-|z|^3)^3$ for $|z|<1$ and 0 otherwise.
\end{itemize}

\begin{figure}[h!]
\footnotesize
\stackunder[5pt]{\includegraphics[width=8cm,height=6cm]{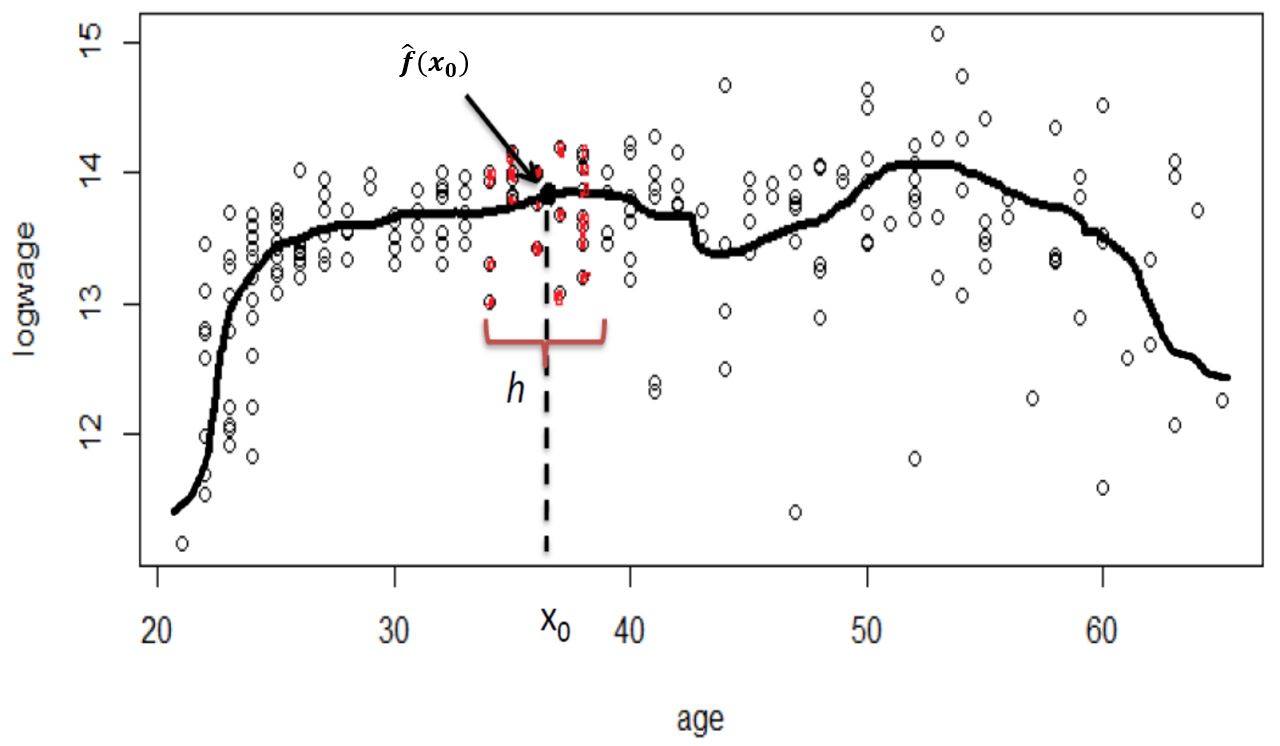}}{(a)}%
\hspace{0.5cm}%
\stackunder[5pt]{\includegraphics[width=8cm,height=6cm]{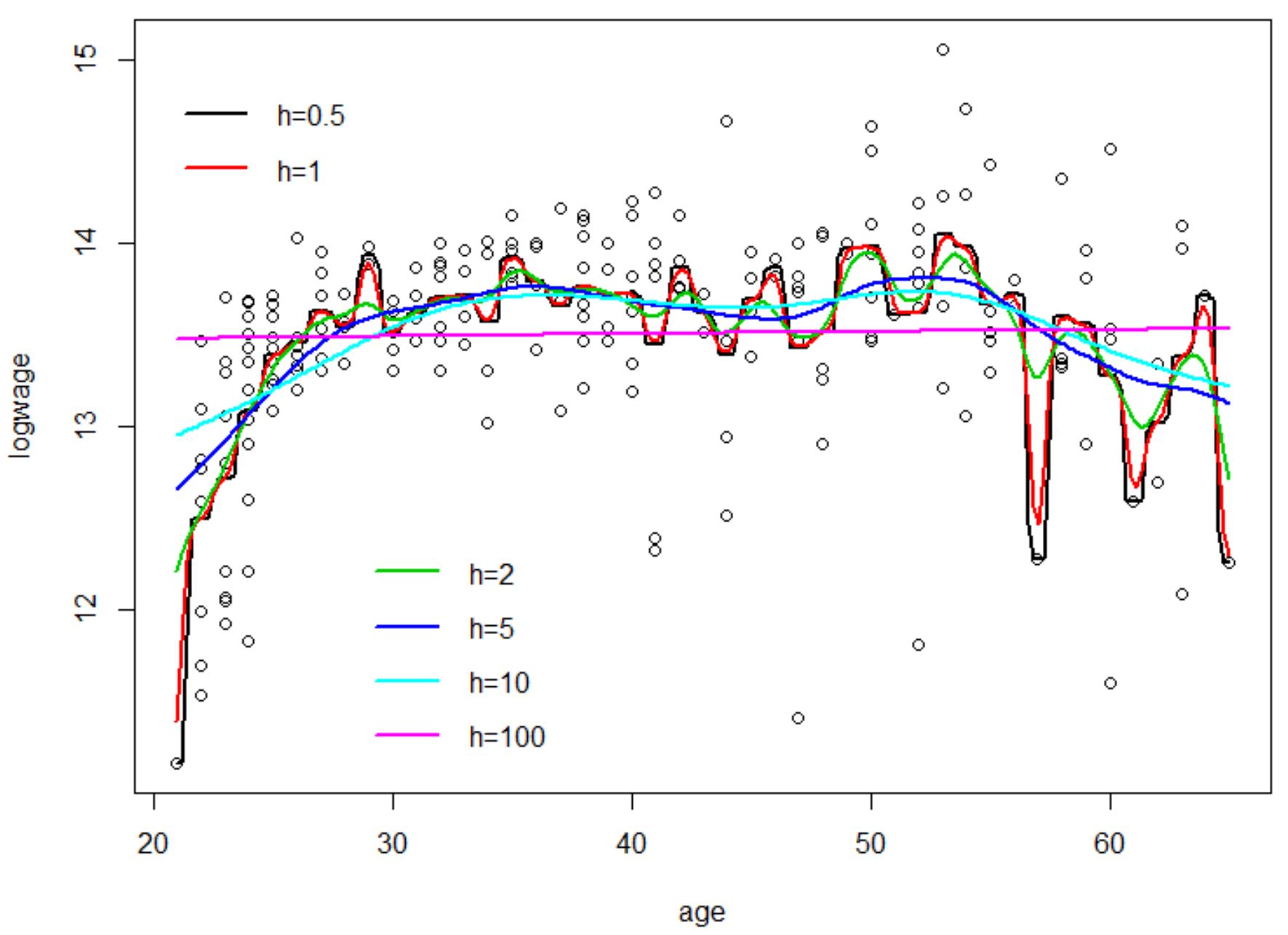}}{(b)}
\caption{Scatter plot and kernel smoothing at $h$ smoothing parameter (a), and estimating the unknown function by Kernel smoothing using $Ksmooth$ function at different value of smoothing parameter, $h$.}
\end{figure}

Bandwidth has an impact on the estimation of the nonparametric function. Figure 5(lest) shows the idea of bandwidth and Figure 5(right) show the effect of the bandwidth on the smoothness of the estimated function. Kernel smoothing of the unknown function and its derivative can be used to see if the curvature is significant or not. Figure 6 shows the smoothed estimated function and its derivative along with 95\% confidence interval using the optimal value of bandwidth. The figure shows the function is increasing and at some point it is a constant. 
\begin{figure}[h!]
\footnotesize
\stackunder[5pt]{\includegraphics[width=8cm,height=6cm]{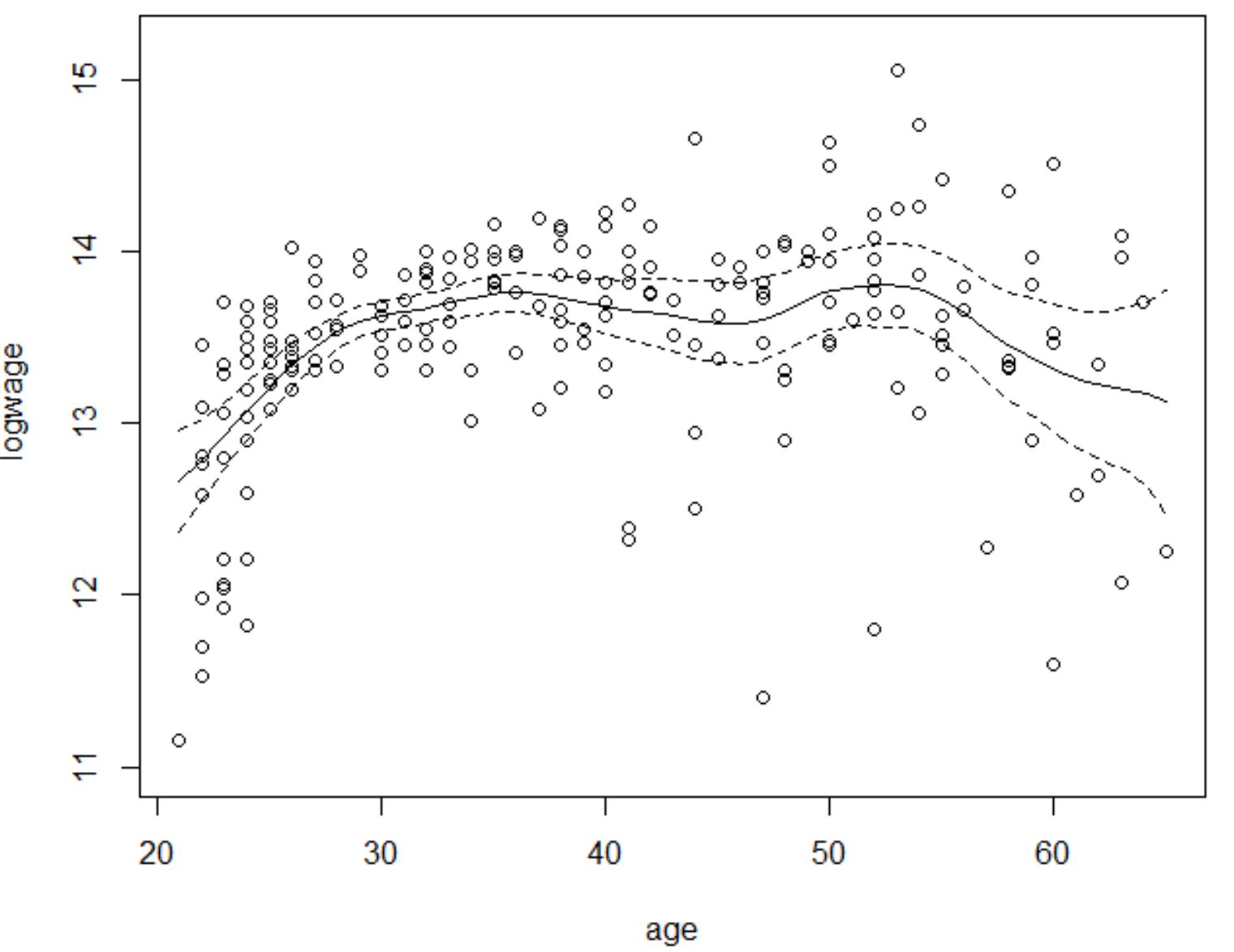}}{(a)}%
\hspace{0.5cm}%
\stackunder[5pt]{\includegraphics[width=8cm,height=6cm]{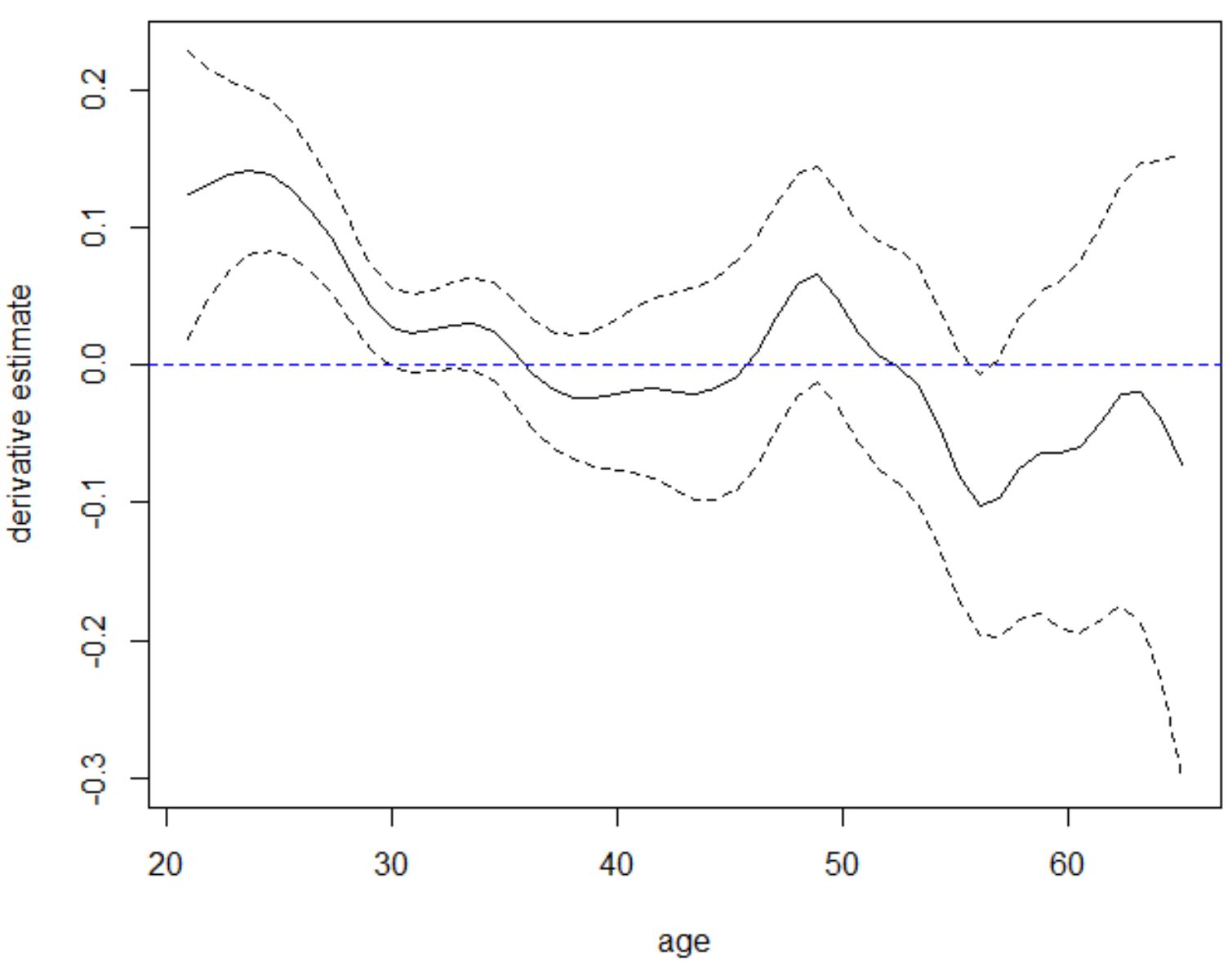}}{(b)}
\caption{Estimate of the unknown function (a), and its derivative with 95\% confidence interval (b) using $npreg$ function at optimal smoothing parameter, $h$, determined my CV criteria.}
\end{figure}
Kernel smoothing does not work well at the boundaries especially if local constant smoothing is used, that is why confidence interval is wide at the boundaries.

\subsection{Spline Smoothing}
Spline is a piecewise polynomial. The polynomials of order $j$ are pieced together at a sequence of knots ($\theta_1 < \theta_2<..... < \theta_C$) such that spline and its first $j-1$ derivative of the spline are continuous at those knots. In spline estimate, the unknown function is approximated by a a power series of degree $p$,
$f(x)=\beta_0+\beta_1x+\beta_2x^2+\beta_3x^3+ ....+\beta_px^p+\sum_{c=1}^C\beta_{1c}(x-\theta_c)_+^p,$\\
where $(x-\theta_k)_+=x-\theta_c, x > \theta_c $ and 0 otherwise.

$\boldsymbol\beta$ is obtained by minimizing the following quantity which has a "roughness penalty", $\lambda$:
\bse
\Sigma_{i=1}^{n}[Y_i-\hat{f}(x_i)]^2+\lambda \int[\hat{f^{''}}(x)]^2dx,
\ese
where $\lambda$ is a tunning parameter controls smoothness and $\hat{f^{''}}(x)$ is the second derivative estimate. There are many methods for choosing $\lambda$, such as CV and GCV (Wahba 1977). For more details in spline smoothing, see Wang (2011).

How many knots need to be used? and where those knots should be located? A possible solution is to consider fitting a spline with knots at every data point, so it could fit perfectly and estimate its parameters by minimizing the usual sum of squares plus a roughness penalty. The smoothing parameter $\lambda$ has the same effect on smoothing such as $h$ in kernel smoothing regression. When $\lambda$ is very big the smoothing ling would be similar to linear regression line and when it is small, the smoothing line would be wiggly. Figure 7(left) shows the smoothed estimated function using optimal $\lambda$ and Figure 7(right) show the effect of the smoothing parameter on the smoothness of the estimated function. Spline smoothing of the unknown function and its derivative can be used to see if the curvature is significant or not. Figure 8 shows the smoothed estimated function and its derivative along with 95\% confidence interval using the optimal value of bandwidth. 

\begin{figure}[h!]
\footnotesize
\stackunder[5pt]{\includegraphics[width=8cm,height=5.5cm]{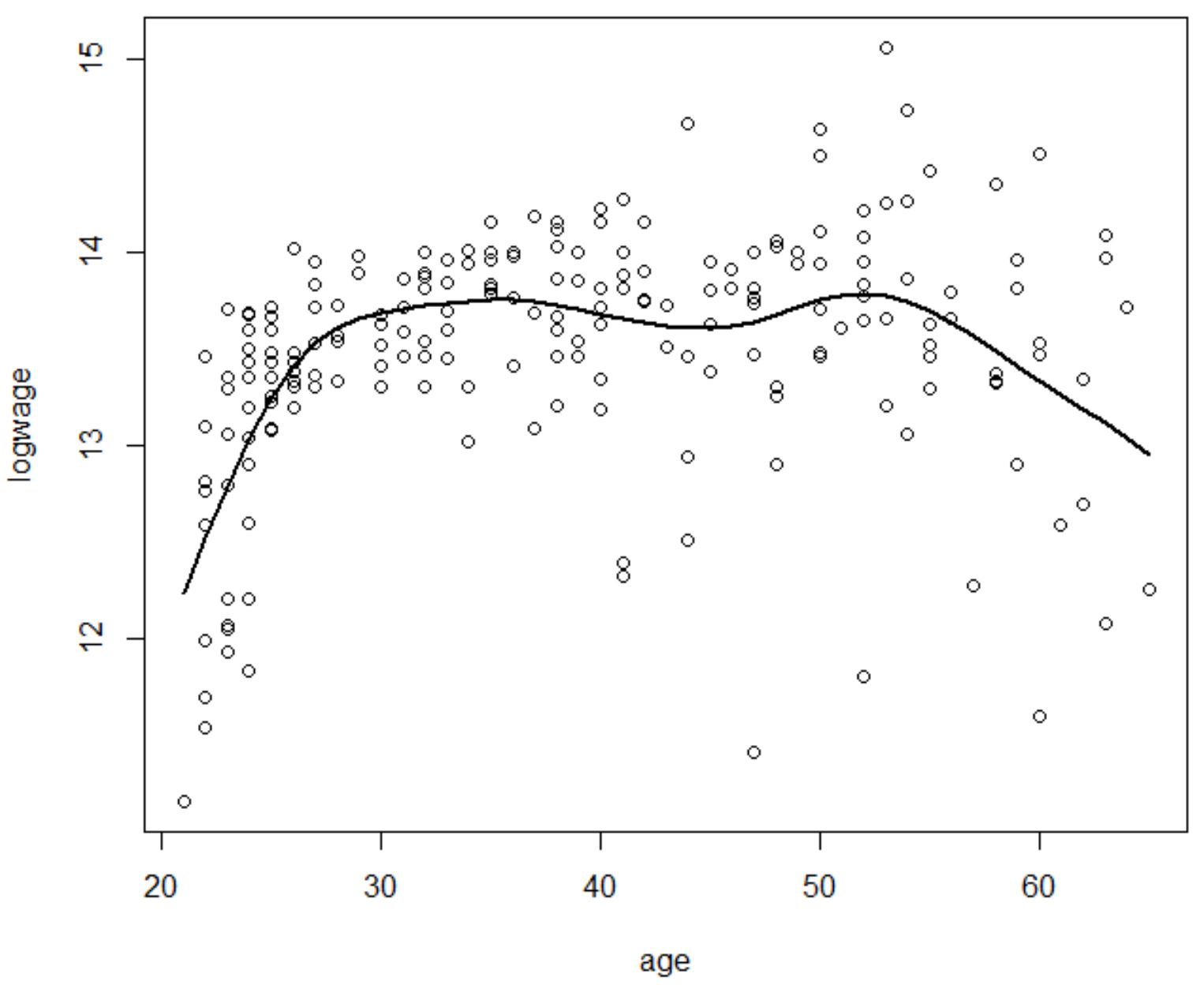}}{(a)}%
\hspace{0.5cm}%
\stackunder[5pt]{\includegraphics[width=8cm,height=5.5cm]{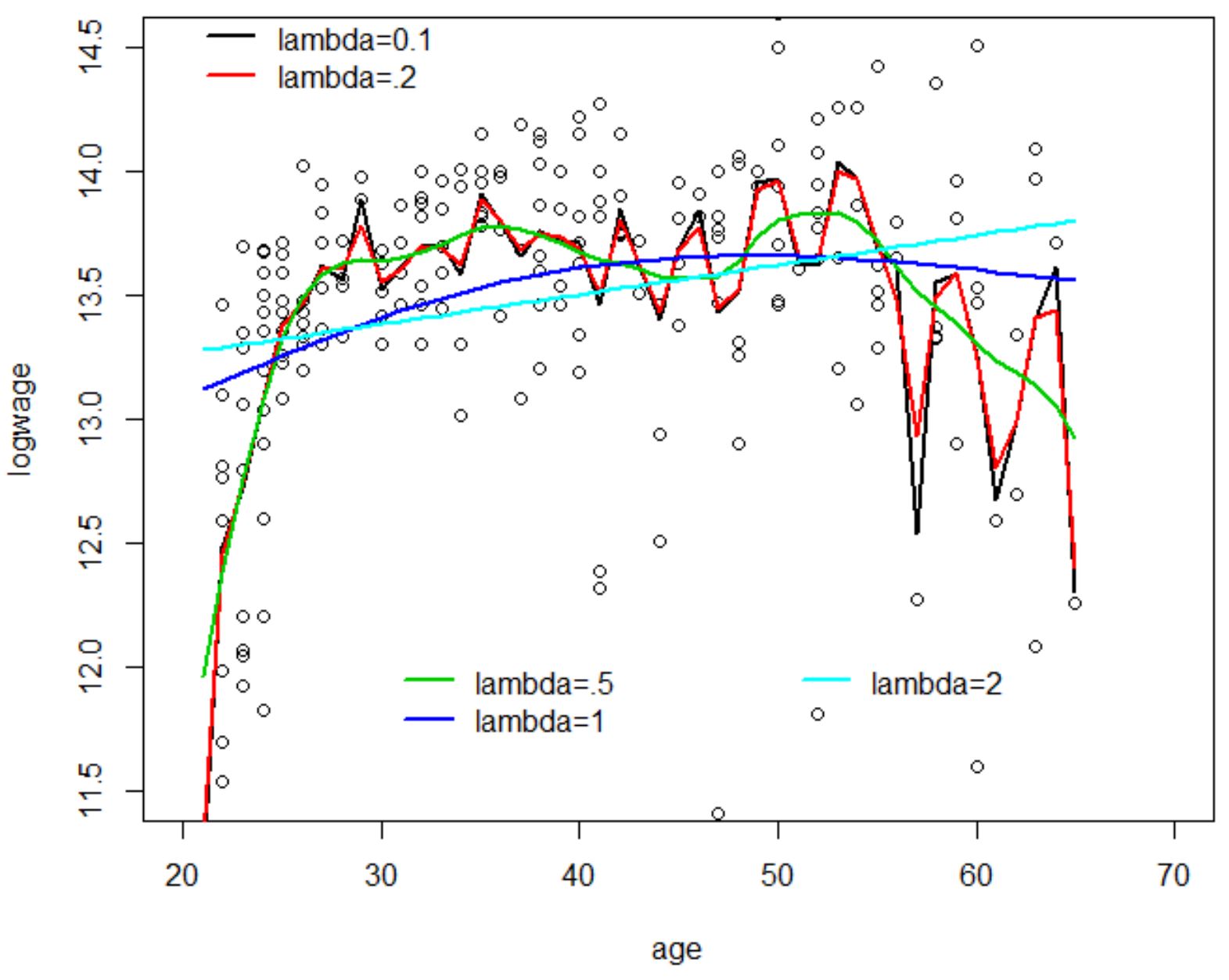}}{(b)}
\caption{Estimate of the unknown function using optimal $\lambda$ determined my GCV criteria (a), and estimation using different values of penalty value, $\lambda$.}
\end{figure}

\begin{figure}[h!]
\footnotesize
\begin{center}
\stackunder[5pt]{\includegraphics[width=8cm,height=4.8cm]{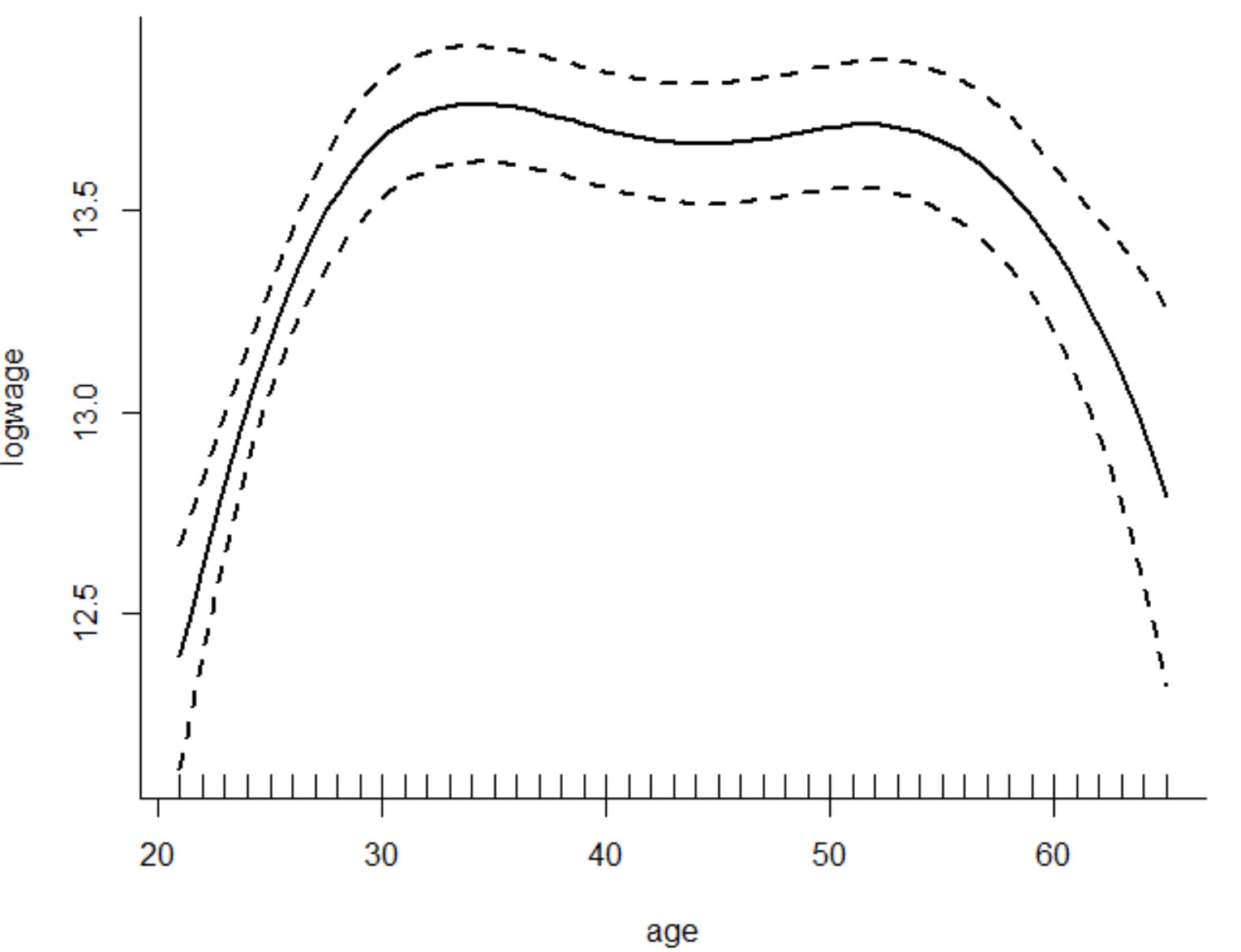}}{(a)}%
\stackunder[5pt]{\includegraphics[width=8cm,height=4.8cm]{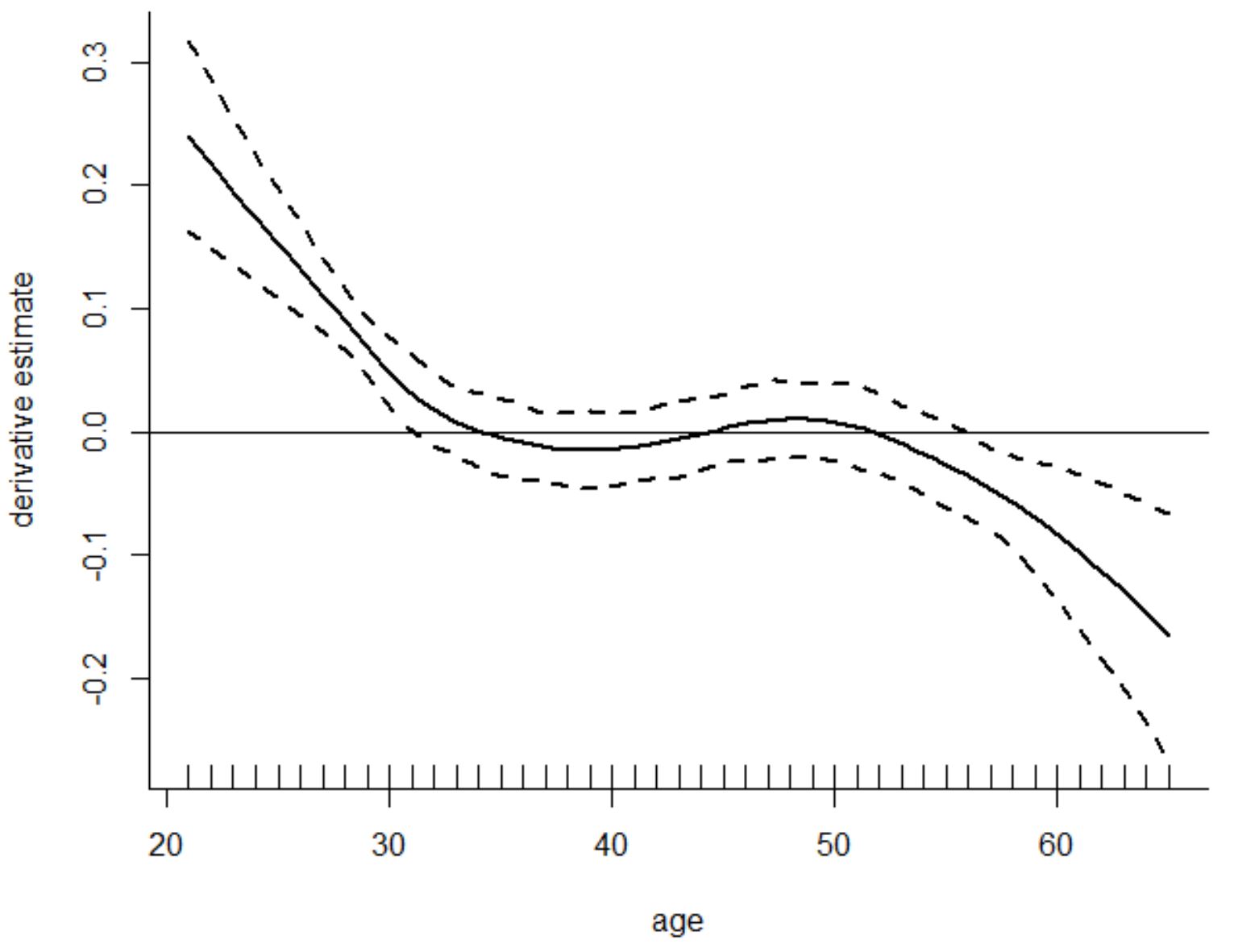}}{(b)}
\end{center}
\caption{Estimate of the unknown function (a), and its derivative with 95\% confidence interval (b) using $spm$ function at optimal smoothing parameter, $\lambda$, determined my REML criteria. Number of knots used is equal to 10 and $\lambda=25.65$.}
\end{figure}

There are many R functions that can be used to estimate semi/nonparametric regression models using kernel and smoothing splines techniques. Table 1 display R packages and functions to estimate the unknown function nonparametrically. 
\begin{table}[h!]\footnotesize
\caption{Different R packages and functions for estimating the nonparametric function and its derivative} 
\centering 
\begin{center}
\begin{tabular}{lllc}
\hline
&Package  & Function&  Smooth parameter\\
\hline
Local Polynomial &locfit  & locfit &  GCV\\ 
Kernel smoothing&KernSmooth & locpoly &  ---\\ 
 &lokern  & glkerns &  plug-in$$\\ 
  &lpridge  & lpridge &  ---\\ 
&np& npreg&\\
&stats& ksmooth&\\
  \hline
 Spline&pspline  & smooth.Pspline &  CV/GCV\\ 
  &sfsmisc  & D1D2 &  GCV\\ 
 &SemiPar  & spm &  RE(ML)\\ 
\hline
\end{tabular}
\label{tab:tab1}
\end{center}
\end{table}

\section{Multiple Case}
These methods, kernel and spline smoothing, can be extended to any number of explanatory variables. Assume a researcher wanted to fit prestige as a response variable with education and income. Figure 1 shows that the relationship between education and prestige is linear and the relationship between income and prestige is nonlinear (unknown). So the following semiparametric model can be assumed
\begin{equation}
Y=\beta_0+\beta_1education+f(income)+\epsilon,
\end{equation}
where $f(\cdot)$ is a nonparametric function and needs to be estimated.
Smoothing splines or kernel regression can be used to fit this the nonparametric term. Figure 9 shows the estimated two functions: a linear function of prestige and education and the nonparametric function of prestige and income. Also the two relationships can be assumed unknown and the following model can be assumed
\begin{equation}
Y=f(education)+f(income)+\epsilon,
\end{equation}
where $f_1(\cdot)$ and $f_1(\cdot)$ are nonparametric functions and need to be estimated.

Figure 10 reveals the estimated nonparametric functions assuming both relationships are unknown, 

\begin{figure}[h!]
\begin{center}
\vskip -1mm
  \includegraphics[width=16cm, totalheight=7cm]{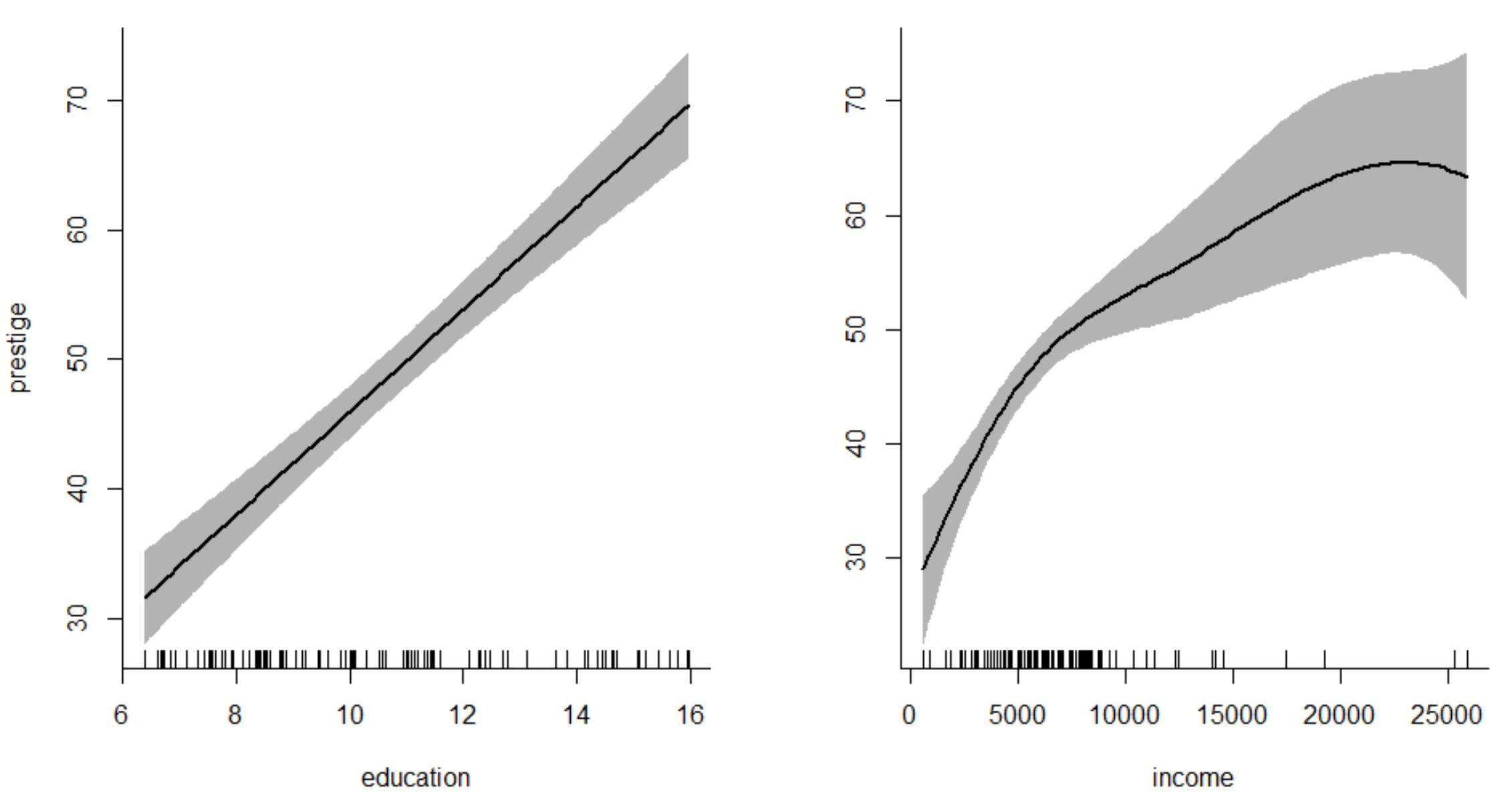}
\end{center}
\vskip -7mm
\caption{Estimated relationship between eduacation, income and prestige using smoothing spline for the model $Y=\beta_0+\beta_1Education + f(Income)+\epsilon$}
\end{figure}

\begin{figure}[h!]
\begin{center}
\vskip -5mm
  \includegraphics[width=16cm, totalheight=7cm]{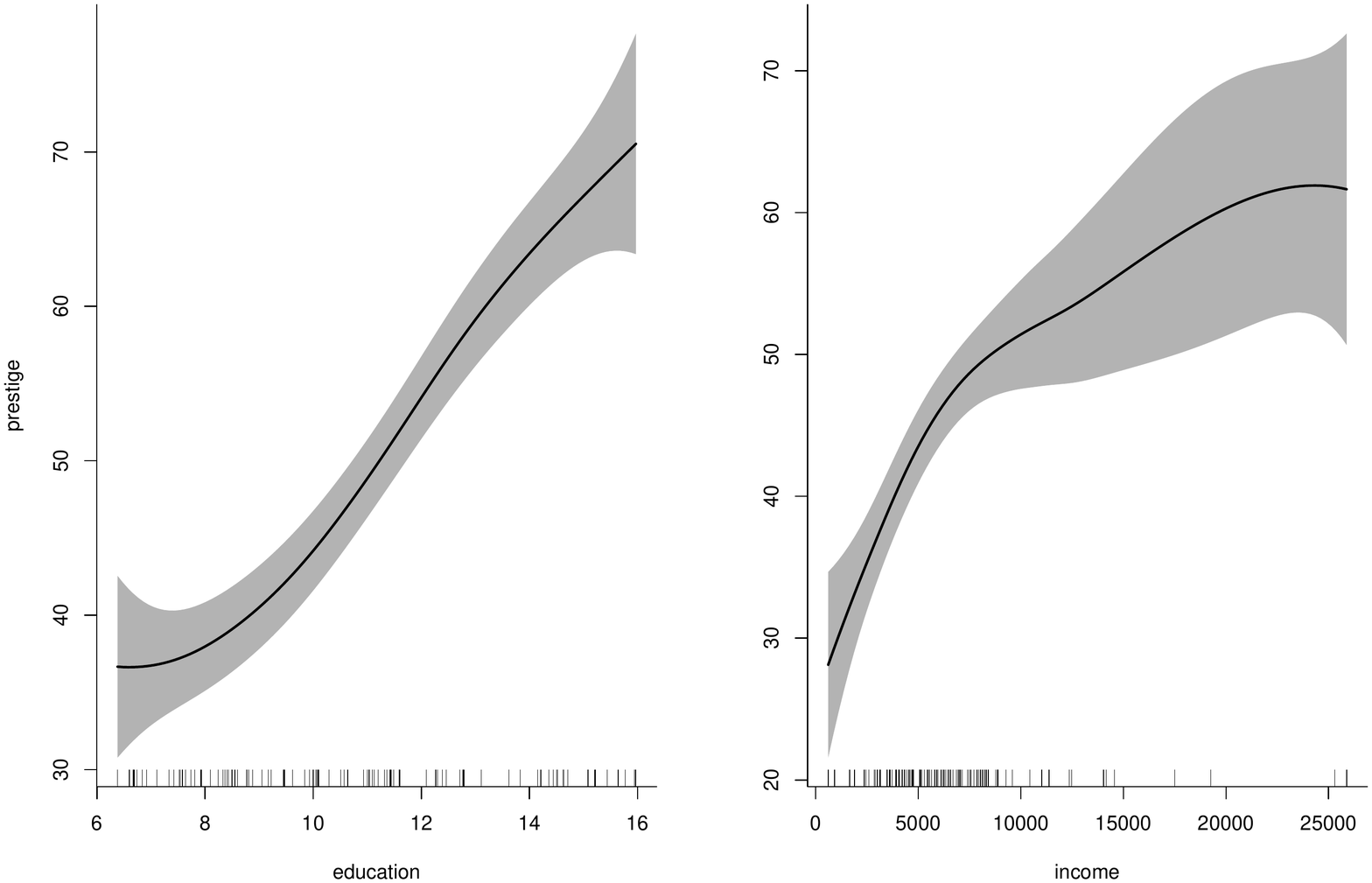}
\end{center}
\vskip -7mm
\caption{Estimated relationship between eduacation, income and prestige using smoothing spline for the model $Y=f(education) + f(income)+\epsilon$}
\end{figure}

Another example for multiple case, an application from Wooldridge (2003, pg. 226) is considered. These data has quantitative and categorical variables and number of observation is 526. It is available in R in $"np"$ R package. The response variable is log(wage)  and the explanatory variables are educ (years of education), exper (the number of years of experience), and tenure (the number of years of current employment), female (Female or Male), and married (Married or Notmarried). The scatterplots are shown in Figure 11 that shows the scatterplots of the explanatory variables with the response. A researcher can assume the relationships between the response and quantitative variables are unknown and estimate them nonparametrically and fit the following model
\begin{center}
$Y=\beta_0+\beta_1 female+f_1(edu)+f_2(exper)+f_3(tenure)+\epsilon$,
\end{center}
where the categorical variable is considered as a factor and the the form of the other relationships, $f_1(\cdot)$, $f_2(\cdot)$, and $f_3(\cdot)$, are unknown.

\begin{figure}[h!]
\begin{center}
  \includegraphics[width=11cm, totalheight=7cm]{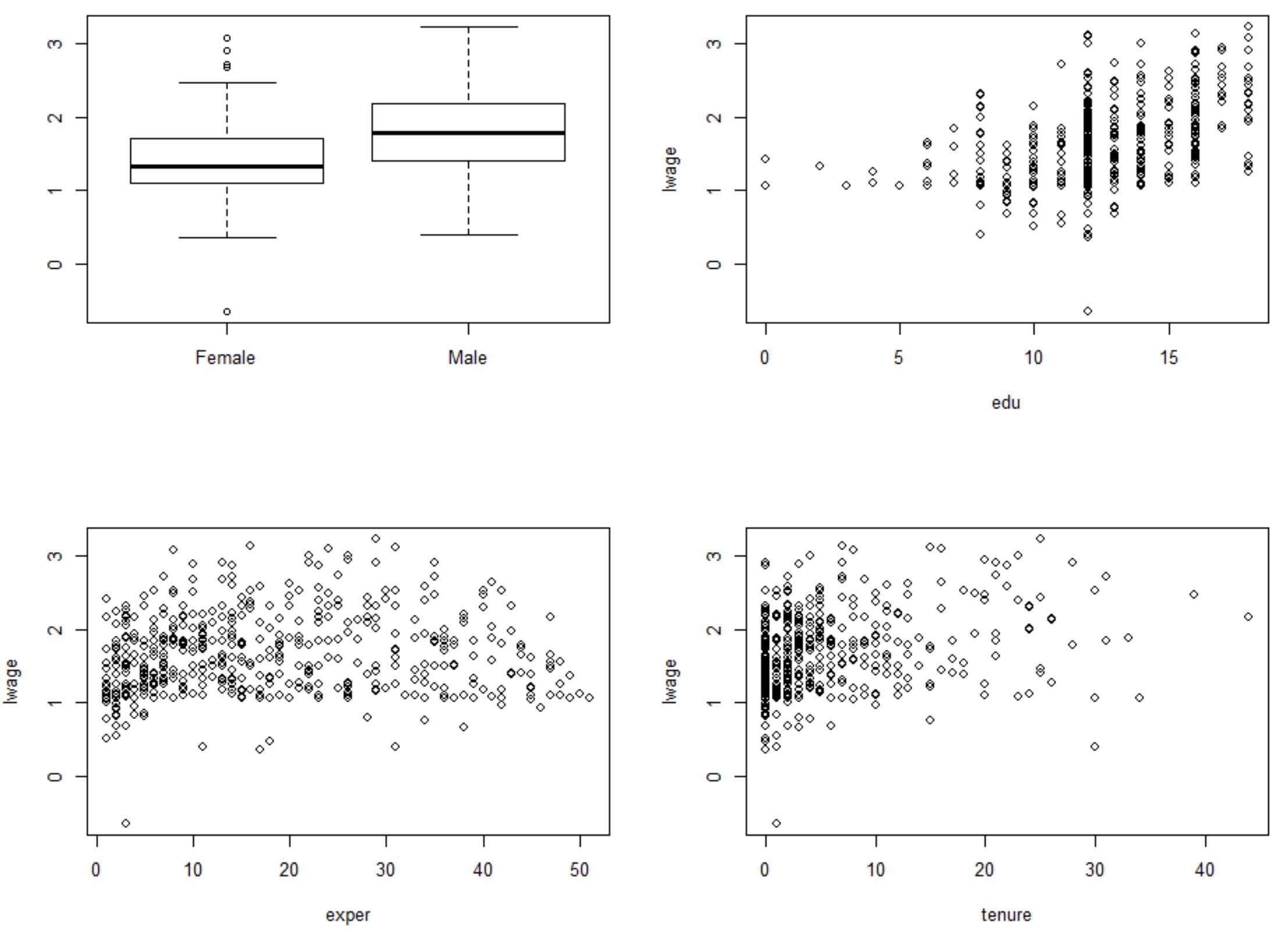}
\end{center}
\vskip -6mm
\caption{Scatter plots of wage data}
\end{figure}
The output are displayed below and the smoothed functions are showed in Figure 12.
\begin{figure}[h!]
  \includegraphics[width=11cm, totalheight=5.5cm]{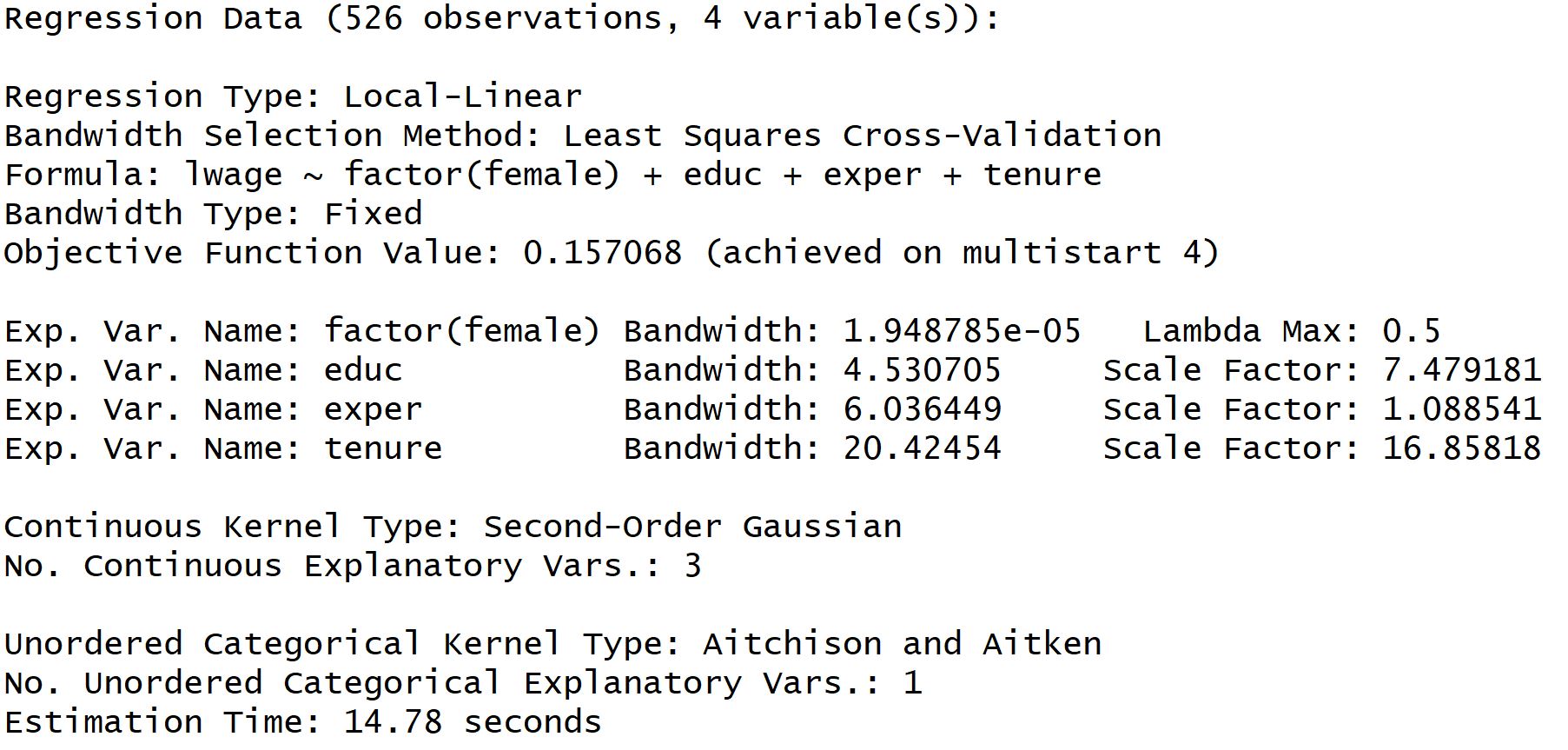}
\end{figure}

\begin{figure}[h!]
\begin{center}
  \includegraphics[width=12cm, totalheight=8cm]{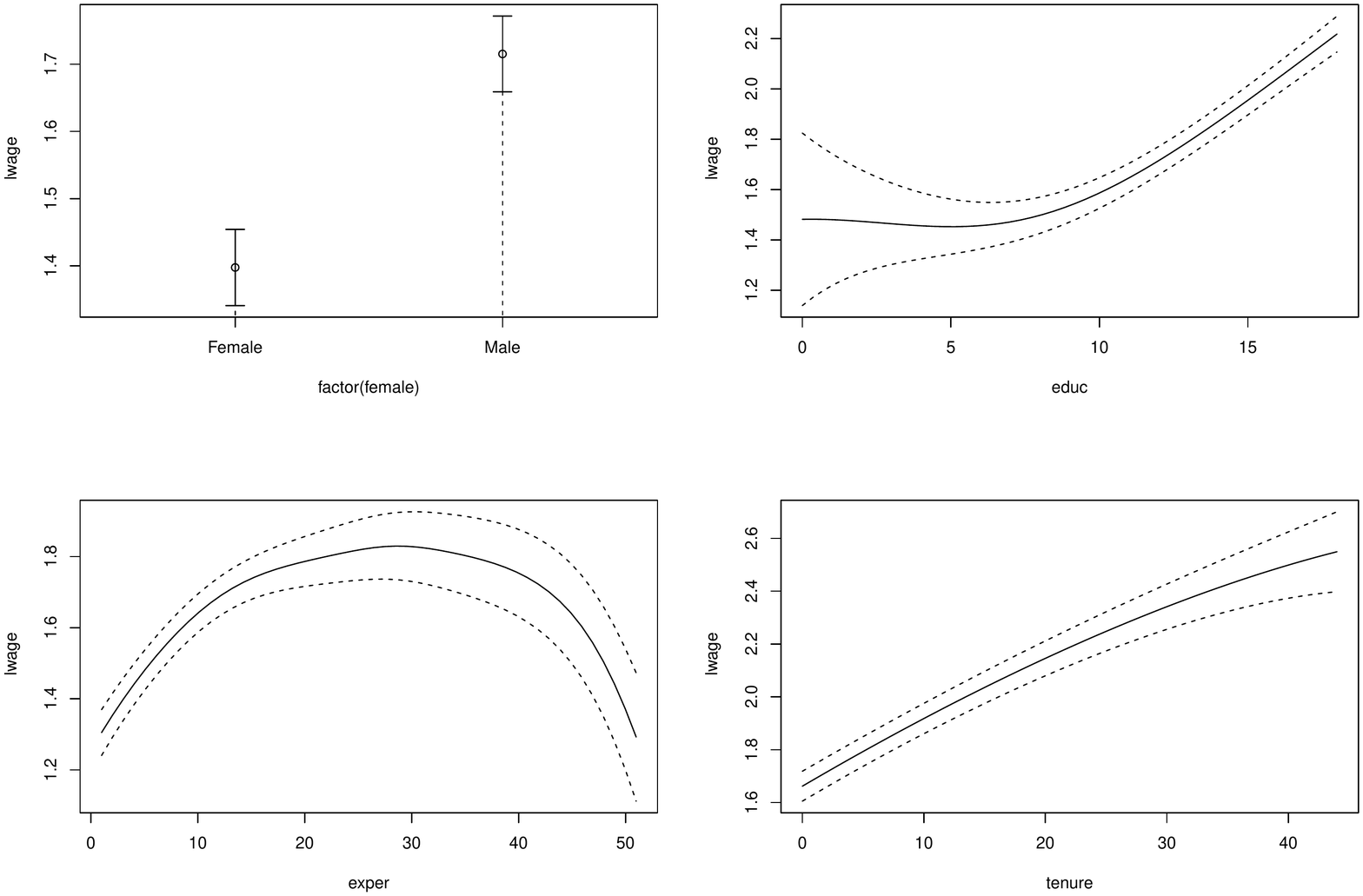}
\end{center}
\vskip -6mm
\caption{Smoothed estimated functions of wage data for the model $Y=\beta_0+\beta_1 female+f(edu)+f(exper)+f(tenure)+\epsilon$}
\end{figure}

Another structure of semiparametric regression modeling, may assumed to this data set, is single index model which takes the following form
\begin{center}
$Y=f(\beta_1tenure+\beta_2female+\beta_3edu+\beta_4exper)+\epsilon$,
\end{center}
where $f(\cdot)$ is unknown function, $\boldsymbol\beta=(\beta_1,\ldots,\beta_4)$ is the single index coefficient vector parameters, and $\epsilon$ is the random error term. The estimated index parameters, R-square, and optimal bandwidth are displayed below and the smoothed fitted function, $\hat{f}(\cdot)$, is showed in Figure 13.

\begin{figure}[h!]
  \includegraphics[width=9cm, totalheight=5cm]{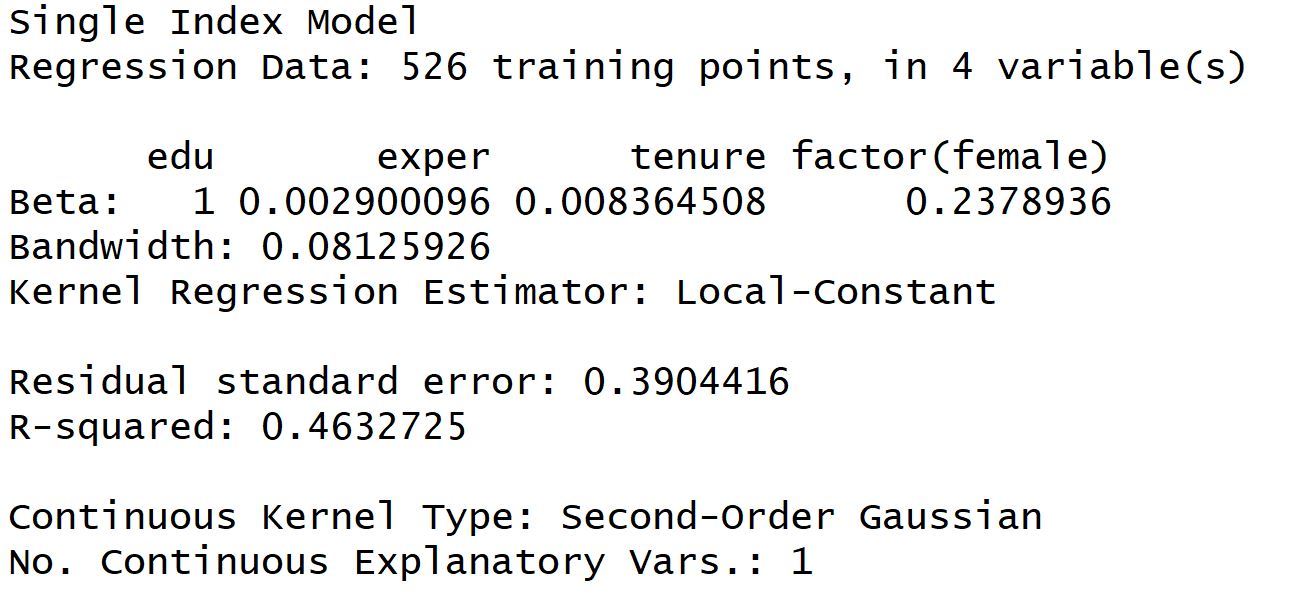}
\end{figure}

\begin{figure}[h!]
\begin{center}
  \includegraphics[width=13cm, totalheight=8cm]{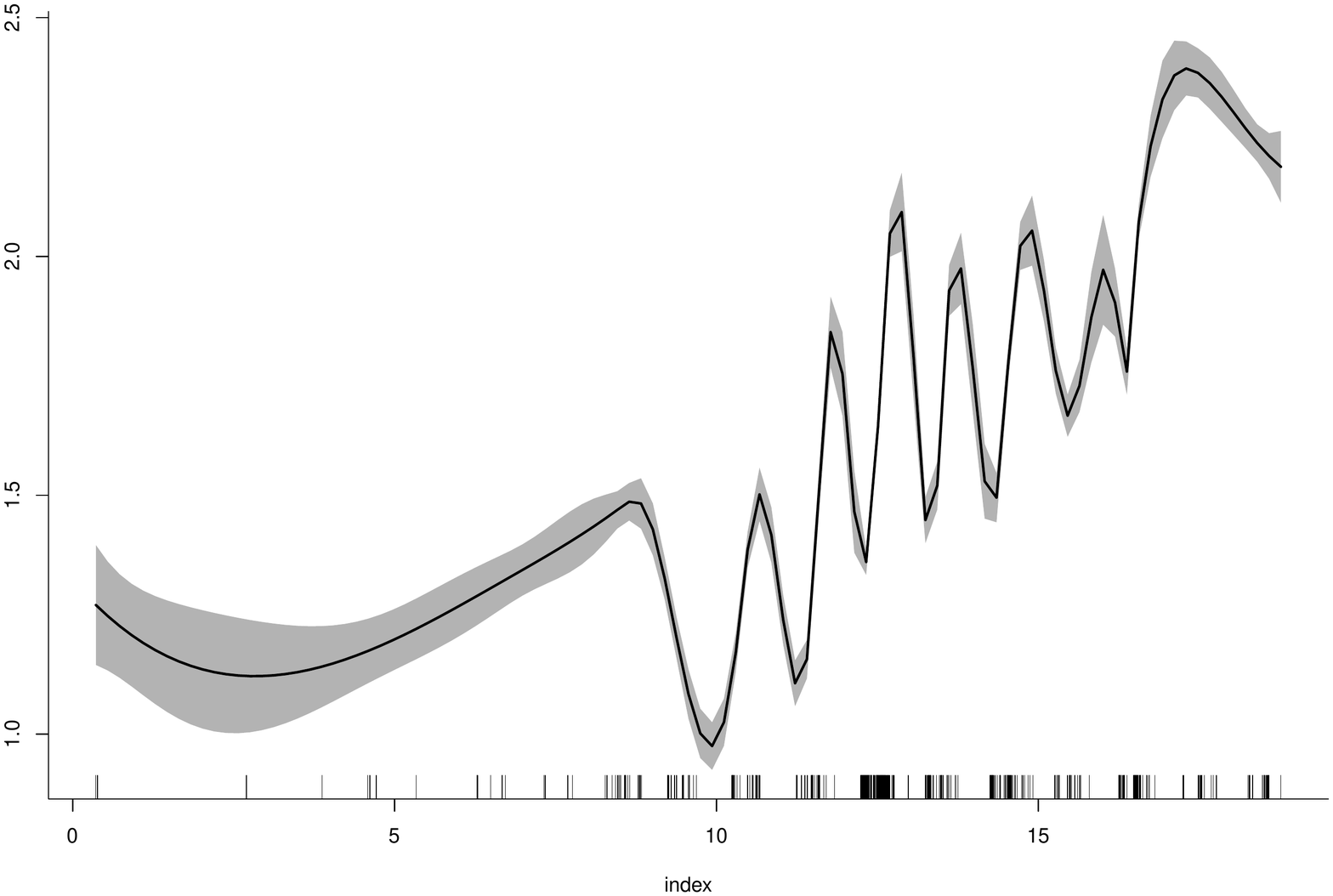}
\end{center}
\vskip -6mm
\caption{Smoothed estimated single index function of wage data for the model $Y=f(\beta_1tenure+\beta_2female+\beta_3edu+\beta_4exper)+\epsilon$}
\end{figure}

\vskip 15cm
\section{Robust Nonparametric Estimation}
Outliers may affect kernel or spline nonparametric estimation. So robust estimation is needed in this case. In kernel smoothing, at a point $x$, the smoothing is obtained from fitting the $p$th-degree polynomial model using weighted least squares with kernel weights as described in equation (7). The kernel function $K$ is usually taken to be a symmetric positive function with $K(d)$ is decreasing as the absolute distance, $|d|$, increases. That is, this polynomial is locally fitted by minimizing the quantity in equation (8). To down weight the effect of unusual values, these weights need to be adjusted to be a function of residuals.

The robust weights are defined as $R_{j}(x)=w^k_j(x)*w^r_j$, $(j=1,2,\ldots,n)$, where $w^r_j$ will be explained below and is obtained from the derivative of some appropriately chosen smooth convex loss function such as Huber or bisquare function. $R_{j}(x)$ is expected to downweight the effect of unusual values on estimation. 

How to obtain $w^r_j$? It is defined as a function of the rescaled residuals to have small weight value for outlying observations.  It takes the form 
\begin{align}\label{eq7}
w_j^r=\frac{\phi^{'}\{[y_j-\hat{f}(x_j)]/\hat{s}\}}{[y_j-\hat{f}(x_j)]/\hat{s}}=\frac{\phi^{'}(r^s_j)}{r^s_ j},\,\, j=1,2,\ldots,n,
\end{align}						
where  $y_j$ is the response value, $\hat{f}(x_j)$ is the estimated mean using locally weighted polynomial regression, $\hat{s}$ is the interquartile range of the residuals, and $r^s_ j=[y_j-\hat{f}(x_j)]/\hat{s}$ is the rescaled residual associated with the value $x_j$. When $x_j$ is an unusual value, $r^{s}_j$ is expected to be large, so the weight $w^{r}_j$ will be small.

To evaluate the performance of the robust weights compared to kernel weights, 100 observations are generated form $f(x)=\{1+e^{-20(x-0.5)}\} ^{(-1)}$ where $x$ is generated from Uniform(0, 1) and two outliers or unusual data points are added, (0.8, 0.6) and (0.75, 0.62), manually to the data. Figure 14 shows the estimated function with the outliers added to the data using kernel, spline and robust weights. Figure 14 shows that kernel and spline are not robust at the interval contain the outliers but the robust estimation does not get affected by these added outliers. They do not get affected only at these two points but also in an interval contain these outlying points.

\begin{figure}[h!]
\begin{center}
  \includegraphics[width=11cm, totalheight=8cm]{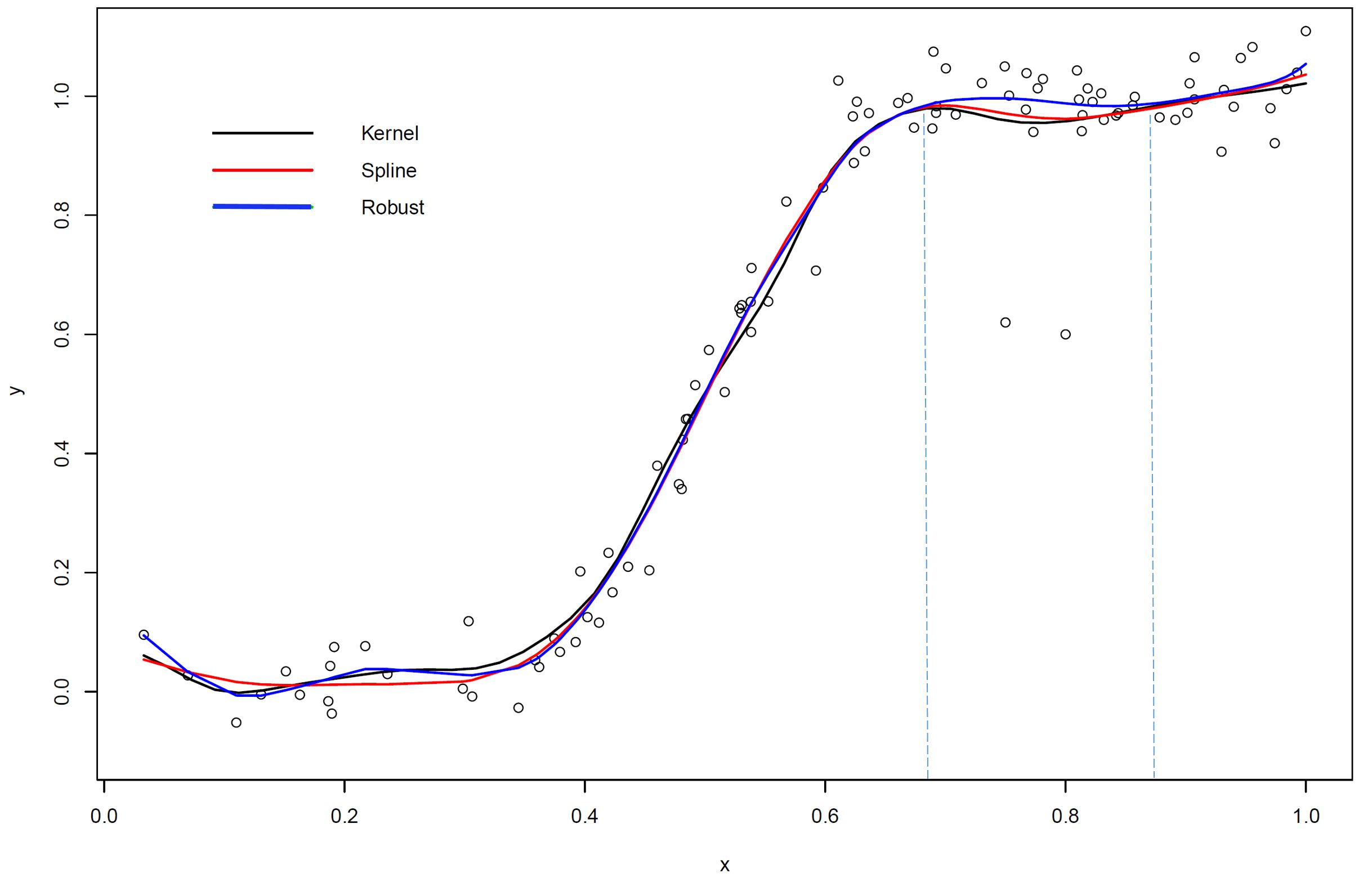}
\end{center}
\vskip -6mm
\caption{Simulated data from $f(x)=\{1+e^{-20(x-0.5)}\} ^{(-1)}$  along with two added outliers points, (0.8, 0.6) and (0.75, 0.62), and the smoothed function by kernel regression, spline and robust nonparametric regression at the same bandwidth, $h=0.046$. }
\end{figure}

\section{Discussion and conclusion}
In this article, three types of modeling are discussed by applications, parametric, semiparametric, nonparametric modeling. Nonparametric and semiparametric regression models are flexible compared to parametric models. Semiparametric is  hybrid of both parametric and nonparametric which allow to have the best of both worlds: a model that is understandable and offering a fair representation of the data in the real life. However, semi/nonparametric regression requires larger sample sizes than regression based on parametric models because the data must supply the model structure as well as the model estimates. semi/nonparametric are flexible and avoid the misleading results when we use a wrong model. One of the well-known semiparametric models is single index model. It assumes the link between the mean response and the explanatory variables is unknown and estimates it nonparametrically. This gives the single index model two main advantages over parametric and nonparametric models: (1) it avoids mispecifying the link function and its misleading results and (2) the reduction of dimension which is achieved by assuming the link function to be a univariate function applied to the projection of explanatory covariate vector on to some direction. Semi/nonparametric is not resistant to outliers, a robust method is introduced. The R code for all the graphs and analyses presented here, in this article, is available in the Appendix.


\vskip 10mm
\noindent
{\Large{\bf References}}
\refmark Fan, J., and Yao, Q. (2003). {\it Nonlinear time series: nonparametric and parametric methods}. Springer: New York.

 \refmark Dhekale, B. S., Sahu, P. K., Vishwajith, K. P., Mishra, P., and Narsimhaiah, L. (2017). Application of parametric and nonparametric regression models for area, production and productivity trends of tea (Camellia sinensis) in India. {\it Indian Journal of Ecology}, 44(2), 192-200.

\refmark Ichimura, H. (1993). Semiparametric least squares (SLS) and weighted SLS estimation of single-index models. {\it Journal of Econometrics}, 58, 71-120.

\refmark Jialiang Li, Chao Huang, Zhub Hongtu, for the Alzheimer’s Disease Neuroimaging Initiative. (2017). A functional varying-coefficient single-index model for functional response data. {\it Journal of the American Statistical Association}, 112:519, 1169-1181

\refmark Lin, W., and Kulasekera, K. B. (2007). Identifiability of single index models and additive index models. {\it Biometrika}, 94, 496-501.

\refmark Loader, C. (1999). Bandwidth selection: classical or plug-in?. {\it The Annals of Statistics}, 27(2), 415-438.

\refmark Mahmoud, H. F. F., Kim, I., and Kim, H. (2016). Semiparametric single index multi change points model with an application of environmental health study on mortality and temperature. {\it Environmetrics}, 27(8), 494-506.

\refmark Mahmoud, H. F. F., and Kim, I. (2019). Semiparametric spatial mixed effects single index models. {\it Computational Statistics \& Data Analysis}, 136, 108-112.

\refmark Nadaraya, E. A. (1964). On estimating regression. {\it Theory of probability and its applications}, 9, 141-142.
 
\refmark Qin. J.,  Yu, T,  Li, P.,  Liu, H.,  and Chen, B. (2018). Using a monotone single index model to stabilize the propensity score in missing data problems and causal inference. {\it Statistics in Medicine}. 38(8) 1442-1458.

 \refmark Rajarathinan, A. and Parmar, R. S. (2011). Application pf parametric and nonparametric regression models for area, production and productivity trends of castor corn. {\it Asian Journal of Applied Sciences}, 4(1), 42-52.

 \refmark Ruppert, D., Wand, M. P., and Carrol, R. J. (2003). {\it Semiparametric regression}.  New York: Cambridge University Press.

 \refmark Wang, Y. (2011). {\it Smoothing splines: methods and applications}. FL: CRC Press, Boca Raton.
 
 \refmark Wand, M. P., and Jones, M.C. (1995). {\it Kernel smoothing}. London; New York: Chapman and Hall.

\refmark Watson, G. S. (1964). Smooth regression analysis. {\it Sankhya, Series A}, 26, 359-372.

\newpage
\vskip 10mm
\noindent
{\Large{\bf Appendix}}

\begin{lstlisting}[language=R]
library(SemiPar); library(np); library(car); library(stats)
library(graphics); data(fossil)                     # Load fossil data 
attach(fossil)
fit = spm(strontium.ratio ~ f(age))
plot(fit)
points(age, strontium.ratio)
plot(age, strontium.ratio)
# ------------------------------------------------------------------
# Dataset cps71 from the np package, which consists of a random sample of 
# wage data for 205 male individuals having common education (grade 13).
data(cps71)
attach(cps71)
str(cps71)                      # gives information about the variables
cps71                           # print the data
plot(age, logwage)
# ------------------------------------------------------------------
# Prestige of Canadian Occupations data set. This data have 6 variables:  
# education, income, women, prestige, census, and type. Source: Canada  
# (1971) Census of Canada. 
data(Prestige)
attach(Prestige)
Prestige
plot(Prestige)
plot(education, prestige, xlab="Education", ylab="Prestige")
# -----------------------------------------------------------------
# ---- Fixing nonlinearity by using polynomial regression 
# -----------------------------------------------------------------
par(mfrow=c(2,2))
plot(age, logwage, xlab="Age", ylab="log(wage)", main="")
linear.model=lm(logwage~age)
lines(age,fitted(linear.model), lty=1, lwd=2, col=4)
plot(linear.model) # check the assumptions
summary(linear.model)
plot(age, logwage, xlab="Age", ylab="log(wage)", main="")
Quad.model <- lm(logwage ~ age + I(age^2))
lines(age,fitted(Quad.model), lty=1, lwd=2, col=4)
plot(Quad.model)
cubic.model <- lm(logwage ~ age + I(age^2)+ I(age^3))
plot(age, logwage, xlab="Age", ylab="log(wage)", main="")
lines(age,fitted(cubic.model), lty=1, lwd=2, col=4)
plot(cubic.model)
Quart.model <- lm(logwage ~ age + I(age^2)+ I(age^3)+ I(age^4))
plot(age, logwage, xlab="Age", ylab="log(wage)", main="")
lines(age,fitted(Quart.model), lty=1, lwd=2, col=7)
plot(Quart.model)
summary(Quart.model)
anova(linear.model, Quart.model)
# -----------------------------------------------------------------
# --------------- First method: Kernel regression 
# -----------------------------------------------------------------
data(cps71)
attach(cps71)
Kernel.smooth <- npreg(logwage~age)
plot(Kernel.smooth, plot.errors.method="asymptotic", 
     plot.errors.style="band" ,ylim=c(11,15.2), 
     main="Estimated function and its 95% confidence band")
points(age,logwage)
summary(Kernel.smooth)
# -------------- Another R function can be used to estimate  
# -------------  the unknown function using kernel regression 
# -------------- Effect of bandwidth on smoothing
plot(age, logwage)
  lines(ksmooth(age, logwage, "normal", bandwidth =.5),  col = 1, lwd=2)
  lines(ksmooth(age, logwage, "normal", bandwidth =1),  col = 2, lwd=2)
  lines(ksmooth(age, logwage, "normal", bandwidth = 2),  col = 3, lwd=2)
  lines(ksmooth(age, logwage, "normal", bandwidth = 5),  col = 4, lwd=2)
  lines(ksmooth(age, logwage, "normal", bandwidth = 10), col = 5, lwd=2)
  lines(ksmooth(age, logwage, "normal", bandwidth = 100),col = 6, lwd=2)
legend(20,15, c("h=0.5", "h=1"),lwd =2, col = 1:2, bty="n")
legend(28,12.5, c("h=2", "h=5", "h=10","h=100"),lwd =2, col = 3:6, bty="n")
# ------------  Smoothing splines
fit <- spm(logwage ~ f(age, degree=3), spar.method="ML")
plot(fit, col=1, shade=FALSE, ylab="logwage")
summary(fit)
plot(fit, se=FALSE)
plot(fit, drv=1, col=1, lwd=2, shade=FALSE, xlab="age", ylab="")
# ------------------------------------------------------------------
# ------------- Second method: Spline regression 
# ------------------------------------------------------------------
# ------ Study the effect of lambda on smoothing
plot(smooth.spline(age,  logwage,  spar = .1), xlab="age", ylab="logwage",
     type="l", xlim=c(20,70), ylim=c(11.5, 14.5), lwd=2)
points(age, logwage)
lines(smooth.spline(age, logwage,  spar = .2), type="l", lwd=2, col=2)
lines(smooth.spline(age, logwage,  spar = .5), type="l", lwd=2, col=3)
lines(smooth.spline(age, logwage,  spar =  1), type="l", lwd=2, col=4)
lines(smooth.spline(age, logwage,  spar =  2), type="l", lwd=2, col=5)
legend(20,14.7, c("lambda=0.1", "lambda=.2"),lwd =2, col = 1:2, bty="n")
legend(30,12.1, c("lambda=.5", "lambda=1"),  lwd =2, col = 3:4, bty="n")
legend(50,12.1, c("lambda=2"),          lwd =2, col = 5, bty="n")

# --------- Fit the function using the optimal soothing value
plot(age, logwage)
lines(smooth.spline(age, logwage), col=1, type="l", lwd=2)

spline=smooth.spline(age, logwage)
plot(spline, type="l", xlim=c(20,70), ylim=c(11, 15), lwd=2)
points(age, logwage)
spline
R2=cor(fitted(spline), logwage); print(R2)
#------------------------------------------------------------------
#---------            Multiple Case
#-------------------- Kernel Regression    ------------------------
data(wage1); wage1
lwage=wage1$lwage; female=wage1$female; married=wage1$married; edu=wage1$edu
exper=wage1$exper; tenure=wage1$tenure
data=data.frame(lwage, female, married, edu, exper, tenure)
par(mfrow=c(2,2))
plot(female, lwage, type="p") ;plot(edu, lwage); plot(exper, lwage)
plot(tenure, lwage)
bw <- npregbw(lwage ~ factor(female) + educ + 
                   exper + tenure, regtype = "ll", bwmethod = "cv.ls",
              data = wage1, tol = 0.1, ftol = 0.1)
npplot(bw, plot.errors.method = "asymptotic", common.scale = FALSE)
summary(bw)
# --------------------- nonparamtric Smoothing Spline  ----------
data(Prestige)
attach(Prestige)
library(mgcv)
plot(income, prestige)
plot(education, prestige)
mod.gam <- gam(prestige ~ s(income) + s(education))
mod.gam
plot(mod.gam)
# ----------------------- Single index model --------------------
bw <- npindexbw(formula=lwage ~ edu+ exper +tenure + factor(female)
                , data=data)
model.index <- npindex(bws=bw, gradients=TRUE)
summary(model.index)
index=model.index$index; est=fitted(model.index)
plot(model.index)
plot(index,est)
fit=spm(est ~ f(index), spar.method="REML")
plot(fit)
# ---------------------------------------------------------------
data(Prestige)
attach(Prestige)
fit1=spm(prestige ~ education + f(income)) # using spline smoothing
fit1=spm(prestige ~ f(education) + f(income)) # using spline smoothing
plot(fit1)
par(mfrow=c(1,2))
plot(fit1, ylab="prestige")

fit2 = npindex(prestige ~ education + income, gradients=TRUE, data =Prestige)
plot(fit2, plot.errors.method="bootstrap",
     plot.errors.center="bias-corrected",
     plot.errors.type="quantiles")
# ---------------------------------------------------------
#------------- Robust Derivative Estimation technique 
x=runif(100,0,1); y=(1+exp(-20*(x-0.5)))^-1 +rnorm(100,0,0.05)
ydash=-(1+exp(-20*(x-0.5)))^-2*(-20)*exp(-20*(x-0.5))
ss=data.frame(x, y, ydash)
newdata <- ss[order(ss[,1]),]
x=newdata[,1]; y=newdata[,2]; ydash=newdata[,3]
x[99]=0.8; y[99]=0.6 ; x[98]=0.75; y[98]=0.62
ss=data.frame(x, y)
newdata <- ss[order(ss[,1]),]
x=newdata[,1]; y=newdata[,2]
# ------------- Kernel smoothing get affected by outliers
Kernel.smooth <- npreg(y~x)
plot(Kernel.smooth, main="", ylim=c(-0.1,1.1))
points(x,y)
summary(Kernel.smooth)
# ------------- Spline soomth also get affected by outliers
cars.spl <- sm.spline(x, y)
lines(cars.spl, col = "red")

# -------------- Robust kernel estimation  
n=length(y)  #sample size
ch=1.345
band=.046
delta=rep(1,n)
whiillr=c(); seyhat=c(); yhat=c(); w=c()
for (ii in 1:8){
    der=c(); est=c()
    for (i in 1:length(x)) 
  {
    u=x-x[i]
    h0j<-exp(-(u/band)^2) 
    sumh<-sum(h0j) 
    h0j<-h0j/sumh
    w0j=h0j*delta
    wxbar=sum(w0j*x)/sum(w0j) 
    wssx=sum(w0j*(x-wxbar)^2)
    w0jllr=(w0j/sum(w0j)) + ((x[i]-wxbar)*w0j*(x-wxbar))/wssx 
    whiillr[i]=w0jllr[i]      
    seyhat[i]=sum(w0jllr^2)   
    w[i]=delta[i]             
    one=rep(1, length(x))
    X=cbind(one,u)
    W=diag(w0j)
    beta=solve(t(X)\%*\%W\%*\%X)\%*\%t(X)\%*\%W\%*\%y
    est[i]=beta[1]
    der[i]=beta[2] 
  }
  yhat=est
  r=y-yhat
  m=median(r)  
  mad=(median(abs(r-m)))/.6745   #mad is the mad estimate of scale
  rs=r/mad                       #rs are the rescaled residuals
  for (j in 1:n){
    if (abs(rs[j]) <=  ch){delta[j]=1} else {delta[j]=ch/abs(rs[j])}
  }
  all=cbind(x, y, yhat ,rs ,delta)
  print(all[1:10,])
}
dferror=sum(w)-sum(whiillr)       #error degrees of freedom
sigma2=sum(w*(y-yhat)^2)/dferror  #estimate of sigma squared
tvalue=2                          #invcdf .975 tvalue
seyhat=sqrt(sigma2*seyhat)        #seyhat contains the se(yhat)
lclm=yhat-tvalue*seyhat
uclm=yhat+tvalue*seyhat
points(x, typ="l", yhat, col="blue", lwd=2)
#lines(x,lclm, col="blue", lwd=2)
#lines(x,uclm, col="blue", lwd=2)


\end{lstlisting}
 
\end{document}